\documentclass[12pt]{article}
\usepackage{comment}
\usepackage{textcomp}
\usepackage{amssymb,amsmath,amsfonts,mathrsfs,mathtools,braket}
\usepackage[utf8]{inputenc}
\usepackage{graphicx,longtable,afterpage}
\usepackage{xcolor}
\usepackage{jheppub}
\usepackage{floatrow}
\usepackage[normalem]{ulem}
\newfloatcommand{capbtabbox}{table}[][\FBwidth]

\allowdisplaybreaks

\newmuskip\pFqmuskip

\newcommand*\pFq[6][8]{%
  \begingroup % only local assignments
  \pFqmuskip=#1mu\relax
  \mathchardef\normalcomma=\mathcode`,
  \mathcode`\,=\string"8000
  \begingroup\lccode`\~=`\,
  \lowercase{\endgroup\let~}\pFqcomma
  {}_{#2}F_{#3}{\left(\genfrac..{0pt}{}{#4}{#5}\Big | #6\right)}%
  \endgroup
}
\newcommand{\pFqcomma}{{\normalcomma}\mskip\pFqmuskip}
\newcommand{\al}{\alpha'}
\title{\vspace{-1.cm} 
One-loop mass corrections and decay widths of Type II heavy string states} 

\author{Massimo~Bianchi\;$^{\natural}$, Maurizio~Firrotta\;$^{\flat}$, Lorenzo~Grimaldi\;$^{\natural,\,\S}$}

\affiliation{
$^\natural$  \href{https://www.fisica.uniroma2.it/}{Dipartimento di Fisica, Universit\`a di Roma Tor Vergata} and \href{https://www.roma2.infn.it/en/roma-tor-vergata-unit/}{INFN sezione di Roma Tor Vergata}, Via della Ricerca Scientifica 1, 00133, Roma, Italy\\
$^\flat$ \href{https://phys.fudan.edu.cn/eng/wbout/list.htm}{Department of Physics and Center for Field Theory and Particle Physics, Fudan University}, Shanghai 200438, China\\
$^\S$ \href{https://www.cref.it/}{Centro Ricerche Enrico Fermi}, Via Panisperna 89A, 00184, Roma, Italy $ $ \\
\vspace{0.3cm} $ $

}

\emailAdd{massimo.bianchi@roma2.infn.it, mfirrotta@fudan.edu.cn, lorenzo.grimaldi@roma2.infn.it}

\abstract{We approach a systematic investigation of the one-loop mass corrections to (super-)string massive higher-spin states. While the imaginary part of the relevant amplitudes is finite, being related to the width of the decay of the states into two lower-mass states at tree level, the real part is generally IR-divergent and needs regularization and renormalization. We mostly focus on states of the first Regge trajectory in the NS-NS sector of Type-II string theories. We explicitly derive a closed-form expression for the integral over the insertion point, relying on properties of elliptic functions and lattice sums. We then regularize the IR divergent integral over the modular parameter of the torus, applying the $i\varepsilon$-prescription in string theory. As a result we compute the desired mass corrections up to level $N = 10$ and analyze their behavior at increasing $N$. Finally, we speculate on the existence of mixing among lower-spin states and conjecture that the one-loop mass matrix be governed by some random matrix theory.}

\date{\today}

\begin{document}

\maketitle

%\hypersetup{pageanchor=true}

%\setcounter{tocdepth}{2}

%\toc

\section{Introduction}
The free string spectrum is highly degenerate. The degeneracy at level $N$ grows as $d(N) \approx e^{\beta_H \sqrt{N}}$ up to powers of $N$ \cite{Hagedorn:1965st}. As soon as one turns on interactions, \textit{i.e.}, at $g_s\neq 0$, all  the massive states become unstable with respect to the decay into lower mass states and mix with each other compatibly with Lorentz invariance\footnote{See, {\it e.g.}, \cite{Bianchi:2003wx} for the case of $AdS$.}. Already at one-loop, one expects mass corrections and intricate mixing. This has already been pointed out in earlier pioneering works \cite{Sundborg:1988ai,Amano:1988ht}, where the authors commented on the difficulty of treating the divergence arising in this analysis. In particular, the imaginary part of the one-loop `amplitude' is expected to be finite and to capture the width of a given string state to decay into two lower mass states at tree level. The real part, instead, is infinite due to IR divergences and needs some regularization and renormalization \cite{Marcus:1988vs}. Relying on the $i\varepsilon$-prescription in string theory \cite{Witten:2013pra, Manschot:2024prc, Eberhardt:2023xck}, the task can be achieved at low level $N=2$ \cite{Sen:2016gqt,Stieberger:2023nol} or for special states, \textit{e.g.}, those in the first Regge trajectory (FRT) with maximal spin, that are unique in that they don't mix with other states in perturbation theory thanks to Lorentz invariance.\\

It is tempting to conjecture that one-loop mixing among states (in the same representation of the Lorentz group) is captured by random matrices whose eigenvalues satisfy level repulsion, very much like nuclear resonances \cite{Wigner:1967qdh}. As a first step towards this ambitious goal, we will study the one-loop mass matrix in the NS-NS sector of the Type II superstrings (A or B doesn't make much of a difference) at level $N=3$ and $N=4$ and for FRT states up to level $N=10$.\\

In order to simplify the analysis, we will work in the time-honored DDF approach (after Del Giudice, Di Vecchia, Fubini) \cite{DelGiudice:1971yjh}, recently revived in the context of cosmic strings \cite{Skliros:2011si} and of chaotic behavior of scattering amplitudes \cite{Bianchi:2019ywd, Gross:2021gsj,Bianchi:2022mhs,Firrotta:2023wem,Hashimoto:2022bll, Bianchi:2023uby}. This allows to identify physical states in the bosonic strings and, with some effort, for the NS-NS sector of the superstring \cite{Aldi:2019osr}. As already observed in \cite{Manes:1988gz}, the polarizations of all physical states are transverse, although not all transverse polarizations are physical \cite{Markou:2023ffh,Basile:2024uxn, Pesando:2024lqa}. This will further simplify the analysis. We will heavily rely on properties of elliptic functions, such as Jacobi $\vartheta$ functions, the Weierstrass elliptic $\wp$ function and their relations with free bosonic and fermionic propagators at one-loop, as well as with characters of $SO(2N)$ current algebras and their decomposition into $SU(N){\times} U(1)$ characters. With some effort, one can extend our analysis to the bosonic string, whereby one has to deal with the instability of the perturbative vacuum signaled by the tachyonic ground state, and to the heterotic string, whereby the tachyon, though projected out thanks to level matching, leaves a subtle phase that complicates the treatment.\\

The reason why we are interested in mass corrections and mixing is twofold. On the one hand, string theory was born as a theory of nuclear resonances and one expects similar effects to take place in other backgrounds that give rise to more viable, possibly holographic, string settings for hadrons and strong interactions. On the other hand, Highly Excited Strings (HES) are potential candidates for micro-states of putative black holes in string theory \cite{Firrotta:2024fvi,Firrotta:2022cku,Das:2023xge,DAppollonio:2015oag,Veneziano:2012yj,Veneziano:2004er,Giddings:2007bw} and we expect the resolution of the large degeneracy of the free string states at high level to play a crucial role in the onset of complexity in the string / BH correspondence \cite{Horowitz:1996nw,Damour:1999aw}.\\

The paper is organized as follows.
In Section \ref{vertex operators section}, we briefly discuss the vertex operators of the physical states we consider in our analysis and set the stage for the one-loop computations.
In Section \ref{contractions section}, we describe the general structure of the relevant Wick contractions at one-loop and the sum over spin structures.
In Section \ref{i epsilon section}, we review the $i\varepsilon$-prescription in string theory and discuss how it leads to the correct regularization and renormalization of the IR divergences at one-loop. 
In Section \ref{n=2 section}, we review the one-loop correction to the mass of the states in the first massive level $N=2$ and confirm the impossibility of mixing (due to Lorentz symmetry) and the degeneracy of states due to supersymmetry. 
In Sections \ref{n=3 section} and \ref{n=4 section}, we study one-loop corrections for FRT states in the second and third massive levels $N=3$ and $N=4$, respectively, and discuss possible mixing among lower spin states.
In Section \ref{summary of results}, we describe the general strategy at higher levels, yet within the restricted set of FRT states in the NS-NS sector, and display their mass corrections and decay widths up to level $N=10$ in tables and plots.  
In Section \ref{conclusions}, we draw our conclusions and outline directions for future investigation.
Useful formulae for the elliptic functions, the partition function and the propagators that we will use in the following are reported in Appendix \ref{appendix formulae}. Appendix \ref{appendix worldsheet integral} contains the full analytic computation of the worldsheet integral, whereas Appendices \ref{appendix characters} and \ref{appendix lattice sums} present alternative computation methods for the first massive level and for an arbitrary massive level, respectively.

%\bibliography{string_scattering}

\section{Vertex Operators and First Regge Trajectory states}
\label{vertex operators section}

Since we are interested in studying one-loop mass corrections and mixing among massive states in Type II superstrings, we need to identify the relevant vertex operators and write down the corresponding one-loop amplitudes. While vertex operators for massless states are very well known, vertex operators for massive string states are much more involved \cite{Bianchi:2010es}. In the covariant approach the conditions for BRST invariance become increasingly more complicated to impose and, except for the first Regge trajectory (FRT), the task becomes prohibitive pretty soon\footnote{Some significant progress in the construction of entire Regge trajectories that has recently been made relies on the observation that the (super-)Virasoro constraints can be written as linear combinations of lowering operators of a bigger (ortho-) symplectic algebra \cite{Markou:2023ffh,Basile:2024uxn}.}.\\ 

One possible way out is to rely on the DDF approach \cite{DelGiudice:1971yjh,Bianchi:2019ywd}, whereby a massive (open bosonic string) state is realized as an excitation of the (tachyonic) ground state with momentum $p_T$ by subsequent emission/absorption of collinear massless photons with null momentum $q$ in such a way that $p_T -N q = p_N=P$ be on-shell, {\it i.e.}, $\alpha' p_N^2 = - \alpha' M_N^2 = 1-N$ that follows requiring  $2\alpha' p_T{\cdot q} =1$.
Denoting as usual the (transverse) bosonic string coordinates by $X^i$ ($i=1,...,D{-}2$), Arbitrary Excited States (AES) can be generated by the action of DDF operators
\begin{equation}
    A^i_{-n} = \oint_{{\cal C}_{\{0\}}} {dz\over 2\pi} \,\partial X^i e^{-in q{\cdot} X}(z)
\end{equation}
that are transverse to the momentum of the state, and the polarization vector can be covariantized as $\zeta^\mu = \lambda^i (\delta_i^\mu{-}2 \alpha' q_i p_T^\mu)$ so that $\zeta{\cdot}P =0$. Manes and Vozmediano have shown that all physical states are transverse \cite{Manes:1988gz}. The reverse is false in general: not all transverse states are physical, {\it i.e.}, BRST invariant, yet all states built with DDF operators are physical.\\
For closed strings, one has to duplicate the procedure in order to account for Left- and Right-movers.  Moreover, using level matching $N=\overline{N}$ and $P_L=P_R=P/2$ the mass-shell condition turns out to be $\alpha' P^2=-\alpha' M^2= 2(2-N-\overline{N}) = 4-4N$.\\
For superstrings, the very definition of DDF operators is extremely involved and cumbersome \cite{Aldi:2019osr}. In the following we will restrict our attention to a simple class of states in the NS-NS sector of Type II superstrings belonging to the first Regge trajectory (FRT), with maximal spin after combining Left- and Right-movers. Later on, we will comment on the generalization to other (highly) excited states in Type II superstring theory and on how to adapt our analysis to heterotic and bosonic strings.\\

As already mentioned, FRT states form a special class of BRST invariant massive ($N\ge 2$) vertex operators with maximal spin $S=N+\overline{N}=2N$ at any given level.  Barring an overall normalization, we find it convenient to write vertex operators for the FRT in the `canonical' (-1) picture of the open (Type I) super-string or for the Left-moving sector of closed (Type II) super-strings in the form
\begin{equation}
{\cal V}^{{\rm super}(-1)}_{FRT}  = e^{-\varphi} \zeta_1{\cdot}\Psi (\zeta_1{\cdot}\partial X)^{N-1} e^{iP{\cdot}X}
\end{equation} 
with $\varphi$ the scalar bosonizing the super-ghost ($\beta, \gamma$) system, $\Psi^\mu$ the fermionic coordinates and $\zeta_1$ a null complex and transverse polarization $\zeta_1{\cdot}\zeta_1 =0=\zeta_1{\cdot} P$. Later, one can replace $\zeta_1^{\mu_1} ... \zeta_1^{\mu_N}$ with $H^{\mu_1\mu_2 ...\mu_N}$ totally symmetric, traceless, and transverse w.r.t. $P$. 
Applying the Picture Changing Operator \cite{Friedan:1985ge} $\Gamma_{+1} = e^{+\varphi} \Psi{\cdot} \partial X + \dots$, the vertex operator in the non-canonical $0$-picture is
\begin{equation}
{\cal V}^{{\rm super} (0)}_{FRT} =  \left[(\zeta_1{\cdot}\partial X)^{N}{+}c_1 P{\cdot}\Psi \zeta_1{\cdot}\Psi (\zeta_1{\cdot}\partial X)^{N-1}
{+}c_2 \zeta_1{\cdot}\Psi\zeta_1{\cdot}\partial\Psi (\zeta_1{\cdot}\partial X)^{N-2}\right]e^{iP{\cdot}X}
\end{equation}
that is needed at one loop in the NS sector, with $c_1 = -i$ and $c_2 = N{-}1$.\\

For Type II superstrings, the full vertex operators take the form 
\begin{equation}
{\cal W}^{(0,0)}_{FRT,2N} = {\cal V}^{L(0)}_{FRT, N} {\cal V}^{R(0)}_{FRT, N}    
\end{equation}
with the Right-moving part similar to the Left-moving one and $\zeta^R_1= \zeta^L_1$, yielding a totally symmetric, transverse, traceless tensor $H_{FRT}^{\mu_1\mu_2 ...\mu_N\tilde\mu_1...\tilde\mu_N}$ of rank $2N=S$.\\ 

Let us discuss the correct normalization of the vertex operators for massive higher-spin states in the FRT. For open (super-)strings one has to impose the correct coupling with the gauge bosons {\it viz.} 
\begin{equation}
{\cal A}_{HHA} = \langle V_H(1) V_H(2) V_A(3)\rangle= {1\over 2} g_{YM} A{\cdot}(p_1-p_2) {1\over N!} (H_1 H_2) + \ldots,
\end{equation} 
where the factors $2$ and $N!$ in the denominator account for the symmetries of $N$ contracted indices in $(H_1 H_2)$ of real spin-$N$ fields (corresponding to FRT states). Since $g_{YM}= g_a (\alpha')^{(10-4)/4}$ where $g_a$ is the open-string coupling ($g_s=g_a^2$), the correct normalization of the disk is $1/g^2_{YM}$ and each vertex comes with a factor of $g_{YM}$ and a power of $(\alpha')^{-N/2}$ to balance the dimension of the worldsheet fields $X$ or rather $\partial^k X$ and $\Psi$  or $\partial^k \Psi$ appearing in the vertex. Moreover, a factor $1/N!$ is needed to produce the correct normalization of the tri-linear vertex after $N!$ possible contractions, with mild modification for superstrings.\\

For closed (super-)strings, the relevant gravitational coupling of two spin $2N$ fields (in the FRT) should be
\begin{equation}
{\cal M}_{HHh} = \langle W_H(1) W_H(2) W_h(3)\rangle = {1\over 2} \kappa (p_1-p_2){\cdot} h{\cdot}(p_1-p_2) {1\over (2N)!} H_1 H_2 + \ldots
\end{equation}
where $\kappa^2 = 8\pi G_N = g_s^2 {\alpha'}^4/(2\pi)^7$. Each vertex comes with a factor of $g_s$ that nicely combines with the topological normalization of the sphere $1/g_s^2$ to produce the correct power of $g_s$ and thus of $\kappa$. Once again, a power of $(\alpha')^{-N}$ is needed to balance the dimension of the worldsheet fields, their derivatives and their Right-moving counterparts.
In addition, a factor $1/N!\sqrt{(2N)!}$ is now needed to produce the correct normalization of the tri-linear vertex after $N!$ contractions. Also in this case, mild modifications are needed for the superstrings.\\

Although we will not deal with them in the present investigation, excited states with lower spin or different representations of the Lorentz group can be built, with the related vertex operator taking the schematic form (in the Left-moving sector)
\begin{equation}
{\cal V}^{L (0)}_{J} = [\zeta_1{\cdot} \partial X \zeta_{1}{\cdot} \partial X + c_1' P{\cdot}\Psi \zeta_1{\cdot} \Psi \zeta_{1}{\cdot}\partial X  +c_2'\zeta_1{\cdot}\Psi\zeta_{1}{\cdot}\partial\Psi] \prod_k (\zeta_k{\cdot}\partial^k X)^{\ell_k}e^{iP_{L} X},
\end{equation}
with $\sum_k k\ell_k =N{-}2$ and $\zeta_k{\cdot} P= 0$. For later use, $L= \sum_k \ell_k \sim J$ will be called  the `length'. States in the FRT correspond to $\ell_1=N-2$ and all the rest $\ell_{k\neq 1} = 0$.

\section{Fermionic and Bosonic contractions and one-loop amplitudes}
\label{contractions section}
We are now ready to write the one-loop amplitude for the mass correction (and mixing) of massive higher-spin states 
\begin{equation}
\mathcal{A}^{(N)}_{\rm 1-loop}=\sum_{s_1,s_2} C_{s_1}C_{s_2}\int_{\mathcal{F}}\frac{d^2\tau}{\tau_2^2}\,{\cal Z}_{(s_1,s_2)}\left<\int_{T^2}d^2z\,\mathcal{W}(z,\bar{z}) \int_{T^2}d^2w\,\mathcal{W}(w,\bar{w})\right>_{(s_1,s_2)}
\end{equation}
where $s_1, s_2$ denote the spin structures of the torus $T^2$, $\tau$ is its modular parameter, $\mathcal{F} $ is the fundamental region of the moduli space, while $C_s$ and $\bar{C}_s$ implement the GSO projection,  and the Type II partition function reads
\begin{equation}
{\cal Z}_{(s_1,s_2)}= {1\over 4\tau_2^{4}} {\vartheta_{s_1}^4(0) \over \eta^{12}} {\bar\vartheta_{s_2}^4(0)\over \bar\eta^{12}},
\label{partition function}
\end{equation}
where $\eta$ is the Dedekind function and the $\vartheta_s$s are Jacobi theta functions\footnote{See Appendix \ref{appendix formulae} for explicit formulae.}. Thanks to the torus isometry  (Conformal Killing Vectors) the integrand only depends on $z-w$, thus one integration trivially produces a factor of $\tau_2$. Setting $w=\bar{w}=0$, we are left with the integration over $z=x+\tau y = x+\tau_1 y + i \tau_2 y$.
\begin{figure}[h!]
    \centering
    \includegraphics[width=0.5\linewidth]{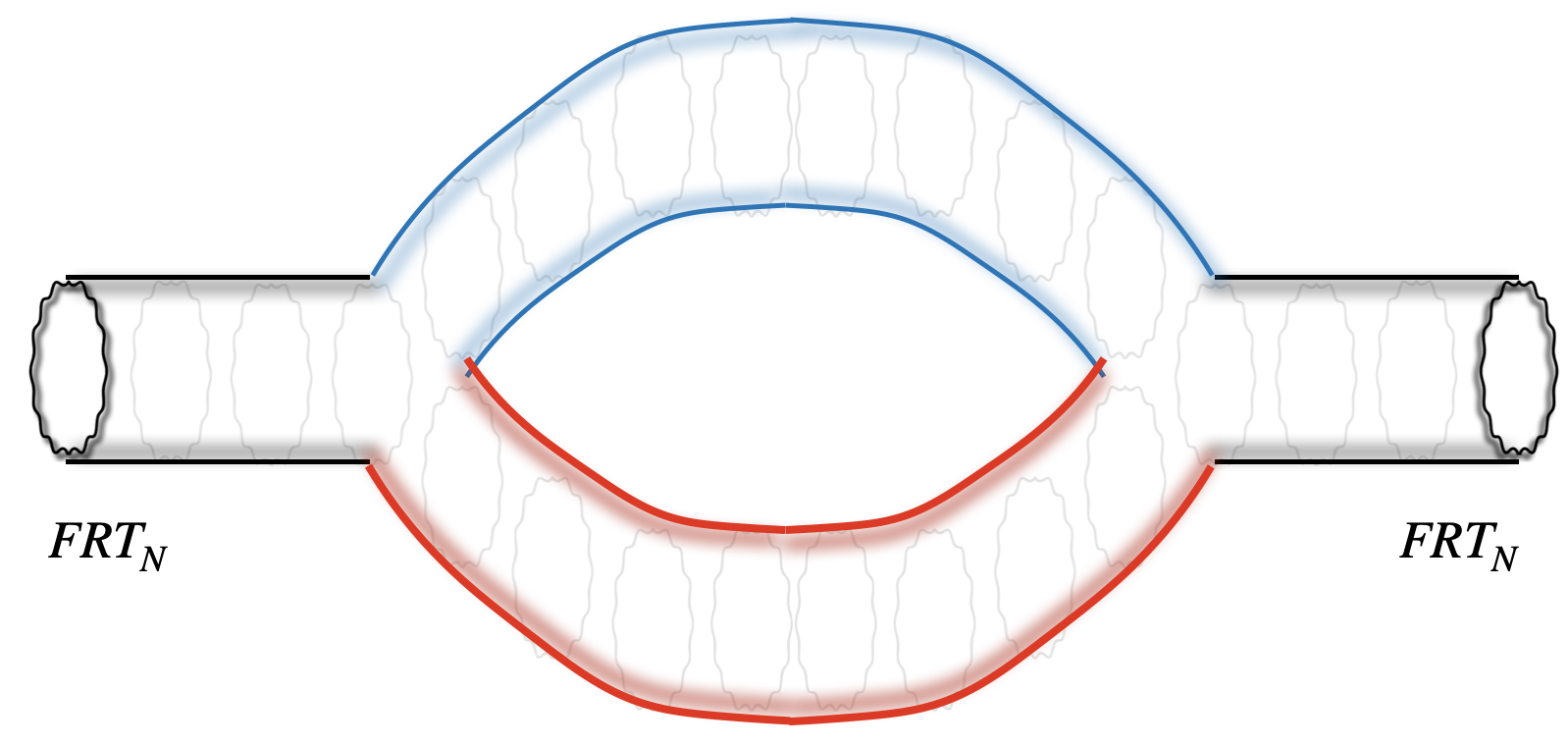}
    \caption{Pictorial representation of the two-point torus amplitude}
    \label{drawing}
\end{figure}

In order to proceed, we must perform the Wick contractions between the bosonic and fermionic coordinates.\\
Setting $\alpha'=2$ henceforth, the bosonic propagator (Bargmann kernel) is
\begin{equation}
\langle X(u_i,\bar{u}_i)X(u_j,\bar{u}_j) \rangle_{{\cal T}}\equiv {G}(u_{ij},\bar{u}_{ij})=-\log \Big| {\vartheta_1(u_{ij} |\tau) \over \vartheta'_1(0|\tau)}   \Big|^{2} + 2\pi{({\rm Im}\,u_{ij})^2 \over  {\rm Im}\,\tau}
\end{equation}
The fermionic propagator (Szego kernel) is 
\begin{equation}
\begin{split}
\langle \Psi(u_i) \Psi(u_j) \rangle_{T^2}^{(s)}=\begin{cases} S_s(u_{ij})= \frac{\vartheta_s(u_{ij}|\tau)}{\vartheta_1(u_{ij}|\tau)}  {\vartheta_1'(0|\tau) \over \vartheta_s(0|\tau)} \,, \quad s=2,3,4  \\ S(u_{ij})=  -\partial_{u_i}\langle X(u_i)X(u_j) \rangle_{T^2} \,, \,\,\,\, s=1\end{cases}
\end{split}
\end{equation}
 with $u_{ij}=u_i-u_j$, and analogously in the anti-holomorphic sector with anti-holomorphic variables.\\
 
The integration of the zero-mode of the bosonic coordinates $x_0^\mu$ results in momentum conservation among contracted vertex operators, that is $P_1=-P_2=P$. Combined with the transversality condition $\lambda_k^{(1)} {\cdot} P_1 = 0= \lambda_k^{(2)} {\cdot} P_2$, we also find $\lambda_k^{(1)} {\cdot} P_2 = 0= \lambda_k^{(2)} {\cdot} P_1$. As a result, all the contractions between the exponentials $e^{\pm iPX}$ and the $\partial^kX$ terms in a generic (yet transverse \cite{Manes:1988gz}) vertex operator vanish.  This is a major simplification. Indeed, the only non-vanishing bosonic contractions are ($z=u_1-u_2$)
\begin{equation}
\langle e^{iP_1X} e^{iP_2X} \rangle {=} e^{-P_1P_2 G(z,\bar{z})} {=} e^{+P^2 G(z,\bar{z})} {=} e^{-M^2 G(z,\bar{z})}  {=} \left\vert {\vartheta_1(z|\tau)\over \vartheta_1'(0|\tau)} \right\vert^{4(N-1)} e^{-4\pi(N-1) {({\rm Im}\,z)^2\over {\rm Im}\,\tau}}
\end{equation}
and the more involved
\begin{align}
    \langle(\partial X)^L(\bar{\partial} X)^L(z) (\partial X)^L(\bar{\partial} X)^L(w)\rangle.
\end{align}
where $L=N-2$ for FRT states. 
The contraction of two primary fields $\partial X$ reads
\begin{align}
    \left\{\begin{aligned}
        &\langle\partial X^i(z) \partial X^j(w)\rangle = \delta^{ij}\partial_z\partial_w G(z-w|\tau)\\
        &\langle\partial X^i(z) \bar{\partial} X^j(w)\rangle = -\delta^{ij}\frac{\pi}{\tau_2}
    \end{aligned}\right.,
\end{align}
from which we see that we also need to consider the `mixed contraction' of holomorphic and anti-holomorphic fields. This is due to the common zero-mode on the torus of the fields $\partial X$ and $\bar\partial X$, that can be identified with the momentum flowing in the loop $2\pi p_0$. In absence of insertions, the integration over $d^Dp_0$ simply produces the factor $(2\pi\alpha' \tau_2)^{-5}$ (in $D=10$), very much as the integration over $d^Dx_0$ produces $(2\pi)^D \delta(\sum_i P_i)$. Instead, if any insertions are present, products of an even number of zero-modes (either from $\partial X$ or from $\bar\partial X$) produce the factor $(2\pi\alpha' \tau_2)^{-5}$ times the `Wick contractions' with Gaussian measure $\exp(-2\pi\alpha' \tau_2 p_0^2)$ 
that yields products of $-\delta^{ij}\frac{\pi}{\tau_2}$. The ones corresponding to the purely holomorphic or purely anti-holomorphic contractions are already included in the Bargmann kernel\footnote{The  purely holomorphic Arakelov kernel $\langle\partial \hat{X}^i(z) \partial \hat{X}^j(w)\rangle = \delta^{ij}\partial_z\partial_w \hat{G}(z-w|\tau)$ does not include them.}. Therefore, in the end, we only have to consider mixed contractions of $\partial X$ with $\bar\partial X$ producing $-\delta^{ij}\frac{\pi}{\tau_2}$, without $\delta^2(z-w)$. Such mixed contractions need to be considered at level $N\geq3$ and were initially neglected in \cite{Grimaldi:2026zgv}.\\

In order to organize the computation, we divide the fields into four sets, namely $A:=\{\partial X(z)\}$, $B:=\{\bar{\partial} X(z)\}$, $C:=\{\partial X(w)\}$ and $D:=\{\bar{\partial} X(w)\}$, where each individual set has cardinality $L$. For later use, it will only be relevant to consider $L=N-2$. The generic term in full Wick contraction will fit in an expansion in the form
\begin{equation}
    \langle(\partial X)^L(\bar{\partial} X)^L(z) (\partial X)^L(\bar{\partial} X)^L(w)\rangle = \sum_{k=0}^L C(k,L)|\partial_z\partial_w G(z{-}w|\tau)|^{2k}\left(\frac{\pi}{\tau_2}\right)^{2(L-k)},
\end{equation}
with multiplicities $C(k,L)$ determined as follows:
\begin{itemize}
    \item Since $k$ contractions are fully holomorphic, we must choose $k$ fields from the set $A$ and $k$ fields from the set $C$. There are $\binom{L}{k}^2$ possible ways to make this choice. In addition, we have to account for the $k!$ permutations of the contracted fields.
    \item We must now contract the remaining $L-k$ elements of $A$ with the same number of elements of $D$, accounting for the mixed contributions. There are $\binom{L}{L-k}=\binom{L}{k}$ possible choices for the subset of $D$ and again we have to consider the $(L-k)!$ possible permutations.
    \item The other $k$ fields in $D$ must be contracted with the same number of fields in $B$, thus giving the fully-antiholomorphic contractions. There are $\binom{L}{k}$ inequivalent choices, with $k!$ permutations.
    \item Finally, the remaining $L-k$ fields in $B$ must be paired with the remaining $L-k$ fields in $C$, building other mixed contributions. At this point, the subset is fixed, but we still have $(L-k)!$ permutations.
    \item Hence, the multiplicity of each term is
    \begin{align}
        C(k,L) = \binom{L}{k}^2 k! \binom{L}{k} (L-k)! \binom{L}{k} k! (L-k)! = (L!)^2 \binom{L}{k}^2.
    \end{align}
\end{itemize}
In conclusion, we have
\begin{flalign}
    \mathrlap{\langle(\partial X)^L(\bar{\partial} X)^L(z) (\partial X)^L(\bar{\partial} X)^L(w)\rangle {=} (L!)^2 {\sum_{k=0}^L} {\binom{L}{k}}^2{|}\partial_z\partial_w G(z{-}w|\tau){|}^{2k}{\left(\frac{\pi}{\tau_2}\right)}^{2(L{-}k)}.}&&
\end{flalign}

%\begin{equation}
% \langle \prod_{k_1} \partial^{k_1} X(u_1)  \prod_{k_2} \partial^{k_2} X (u_2)\rangle  = \sum_{perms}  \prod_\ell \partial^\ell G(z,\bar{z}) 
%\end{equation}
%where $\sum_\ell \ell = 2 \sum_{k_1} k_1 = 2 \sum_{k_2} k_2 $, since $\sum_{k_1} k_1 = \sum_{k_2} k_2$ for any perturbative mixing to take place. Moreover only vertex operators with the same `length' (number of $\partial^k X$ independently of $k$) can contract.  This constraint is obviously satisfied by the states in the FRT with $k=1$ and thus $L=N$.

Schematically, the complete contractions that we must compute read
\begin{align}
\left\langle \prod_{i=1}^2(\partial_iX)^N(\bar{\partial}_i X)^N\right\rangle_s &+ \left\langle \prod_{i=1}^2{P_i}{\cdot}{\Psi} {\Psi} (\partial_i X)^{N-1}\overline{P}_i{\cdot}\overline{\Psi}\,\overline{\Psi}(\bar{\partial}_i X)^{N-1}\right\rangle_s + \nonumber\\
&+ \left\langle \prod_{i=1}^2{\Psi} {\partial_i}{\Psi} (\partial_i X)^{N-2}\overline{\Psi} {\bar{\partial}_i}\overline{\Psi}(\bar{\partial}_i X)^{N-2}\right\rangle_s.
\label{fermionic part}
\end{align}
After summation over the (even) spin structures, only the latter term yields a non-zero result. Indeed, the first one vanishes thanks to Jacobi identity, whereas the second one produces 
\begin{equation}
\sum_s C_s \vartheta_s(0)^2 \vartheta_s(z)^2 = 2 \vartheta_1^2(0) \vartheta_1^2(z) = 0,
\end{equation}
vanishing due to the Riemann identity
\begin{flalign}
\sum_{s=2}^4 C_s \vartheta_s(z_1) \vartheta_s(z_2) \vartheta_s(z_3) \vartheta_s(z_4) &= 
\vartheta_1({+}{+}{+}{+})\vartheta_1({+}{+}{-}{-}) \vartheta_1({+}{-}{+}{-}) \vartheta_1({+}{-}{-}{+})\nonumber&&\\
&\mathrlap{- \vartheta_1({-}{+}{+}{+})\vartheta_1({+}{-}{+}{+}) \vartheta_1({+}{+}{-}{+}) \vartheta_1({+}{+}{+}{-}),}&&
\end{flalign}
where $C_3=1$, $C_2=C_4=-1$ and
\begin{equation}
\pm\pm\pm\pm := {1\over 2} (\pm z_1 \pm z_2 \pm z_3\pm z_4),
\end{equation}
with $z_1=z_2=0$ and $z_3=z_4=z$.\\
Instead, the third term in eq.~\eqref{fermionic part} produces
$$
\sum_s C_s \langle \Psi \partial_1\Psi (u_1)
\Psi \partial_2\Psi (u_2) \rangle_s \sim 
$$
\begin{equation}
\lim_{(u_3, u_4)\rightarrow (u_1, u_2)}  \partial_1 \partial_2 
\sum_s C_s \left[{\vartheta_s(u_{12}) \vartheta_s(u_{34}) \over \vartheta_1(u_{12}) \vartheta_1(u_{34}) } - {\vartheta_s(u_{14}) \vartheta_s(u_{32}) \over \vartheta_1(u_{14}) \vartheta_1(u_{32}) }\right]{\vartheta_s^2(0)} 
\end{equation}
Terms with derivatives acting on the $\vartheta_1(u_{ij})$ in the denominator yield vanishing contributions, thanks to the Riemann identity. For the remaining terms, we find 
\begin{equation}
\langle \Psi \partial_1\Psi
\Psi \partial_2\Psi \rangle_s \sim  [\vartheta_1'(0)]^4,
\label{surviving fermionic contraction}
\end{equation}
since $\vartheta_1(z)^2$ in the numerator exactly cancels an analogous factor in the denominator. This cancels the $\eta^{12}$ in the denominator of the partition function \eqref{partition function}, thus only contributing an overall constant $(2\pi)^4$.\\

Including the Right-movers in the FRT, we get the one-loop self-energy correction 
\begin{equation}
{\cal M}^{(N){\rm Type II }}_{FRT,FRT} = g_s^2 {C^{(N)}\over \alpha'} H^{(1)} _{2N}{\cdot}H^{(2)}_{2N}\frac{(2\pi)^4}{4}{\cal A}^{(N)} = {1\over (2N)!} H^{(1)} _{2N}{\cdot}H^{(2)}_{2N} 
\delta{M}^2 =\delta{S}^{1-loop}_{eff},
\end{equation}
where
\begin{equation}
\delta{M}^2 = 2 M{\rm Re}\delta{M} + i  2 M{\rm Im}\delta{M}= 2 M({\rm Re}\delta{M} + i  \Gamma) = 2{2\sqrt{N-1}\over \sqrt{\alpha'}} ({\rm Re}\delta{M} + i  \Gamma),
\end{equation}
${\rm Re}\delta{M}$ is the mass-shift and $\Gamma$ is the width and we will mostly focus on
\begin{align}
{\cal A}^{(N)}=&\int_{\cal F} \frac{d^{2}{\tau}}{\tau_{2}^{5}} \sum_{k=0}^{N-2}\binom{N-2}{k}^2\left(\frac{\pi}{\tau_2}\right)^{2(N-2-k)}\cdot\nonumber\\
&\cdot \int_{T^2} d^{2}{z} \, e^{-\frac{4\pi z_{2}^{2}}{\tau_{2}}(N{-}1)}\left|\frac{\vartheta_{1}(z|\tau)}{\vartheta_1'(0|\tau)}\right|^{4(N{-}1)} \left|{\partial_z^{2}{G}(z|\tau)}\right|^{2k}
\label{self-energy correction}
\end{align}
Notice that $\vartheta_{1}(z)$ and $\overline{\vartheta_{1}(z)}$ in the numerator make the integrand vanish at $z=0$.\\

\section{Integration over the worldsheet coordinate}
\label{worlsheet integration section}
We will first focus on the worldsheet integral appearing in eq. \eqref{self-energy correction}:
\begin{equation}
    {\cal I}_{ws}(\tau, \bar\tau) = \int_{T^2} d^{2}{z} \, e^{-\frac{4\pi z_{2}^{2}}{\tau_{2}}(N-1)}\left|\vartheta_{1}(z|\tau)\right|^{4(N-1)} \left|{\partial_z^{2}{G}(z|\tau)}\right|^{2k},
    \label{worldsheet integral}
\end{equation}
where $z=x+\tau y$, the integration measure reads $d^{2}z=\tau_2\,dx\,dy$ and the integration is carried out over the fundamental cell $\{x\in (0,\,1),\, y\in (0,\,1)\}$ of the torus $T^2$. This integral can be performed `exactly' relying on the properties of the Weierstrass elliptic $\wp(z|\tau)$ function, the lattice sum formulae defining the Jacobi $\vartheta_1$ and $\vartheta_2$ functions, and on current algebra techniques.\\

At this stage, it is convenient to use the identity
\begin{equation}
    \partial_{z}^{2}G(z|\tau)= \wp(z|\tau)- {\cal E}(\tau),
    \label{bargmann to wp}
\end{equation}
where we have introduced the quasi-holomorphic modular form 
\begin{equation} {\cal E}(\tau):=4\pi i\partial_{\tau}\log{(\eta(\tau)\sqrt{\tau_{2}})}\end{equation}
Since $\wp(z|\tau)$ is related to the Szego kernel through
\begin{equation}
\wp(z) = S_{s}(z)^2 + e_{s-1}(\tau)  
\end{equation}
where $e_{s-1}(\tau)$ denote the Weierstrass roots
\begin{equation}
 e_{s-1}(\tau) = -4\pi i \partial_\tau \log{\vartheta_s(0, \tau)\over \eta(\tau)}
\end{equation}
choosing $s=2$ for convenience, one has
\begin{equation}
    \wp(z|\tau)=e_1(\tau)+\frac{\vartheta_2(z|\tau)^2\vartheta_1'(0|\tau)^2}{\vartheta_1(z|\tau)^2\vartheta_2(0|\tau)^2},
    \label{wp to e1}
\end{equation}
that can be plugged into eq. \eqref{bargmann to wp} to write
\begin{equation}
    \partial_z^2 G(z|\tau)= \tilde{e}_{1}(\tau) + f_{1}(\tau) \frac{\vartheta_{2}^{2}(z|\tau)}{\vartheta_{1}^{2}(z|\tau)},
\end{equation}
where 
\begin{equation}
\tilde{e}_{1}(\tau) := e_{1}(\tau)-{\cal E}(\tau) \quad , \quad 
f_{1}(\tau):=\left({\vartheta_{1}^{'}(0|\tau)\over \vartheta_{2}(0|\tau)} \right)^{2}
\end{equation}
In this way, the worldsheet integral \eqref{worldsheet integral} contains only a product of powers of $\vartheta_1$, $\vartheta_2$, and their complex conjugates, turning into
\begin{align}
    \mathcal{I}_{ws}{=}\tau_2{\int_0}^{1}dx{\int_0}^{1}dy\, &e^{-4\pi(N{-}1)\tau_{2}y^{2}} \left|\vartheta_{1}(x{+}\tau y|\tau)\right|^{4(N{-}1)} \left|\tilde{e}_{1}(\tau){+}f_{1}(\tau)\frac{\vartheta^{2}_{2}(x{+}\tau y|\tau)}{\vartheta_{1}^{2}(x{+}\tau y|\tau)}\right|^{2k}.
    \label{powers of thetas}
\end{align}
Separating the binomial expansions in the holomorphic and anti-holomorphic sectors yields
\begin{flalign}
    &\mathcal{I}_{ws}=\tau_2{\int}_0^1 dx{\int}_0^1 dy\, e^{-4\pi(N-1)\tau_2 y^2} \sum_{n,\,\bar{n}=0}^{k}\binom{k}{n}\binom{k}{\bar{n}} \tilde{e}_1 (\tau)^n \overline{\tilde{e}_1 (\tau)}^{\bar{n}} f_1 (\tau)^{k-n} \overline{f_1 (\tau)}^{k-\bar{n}}\cdot&&\nonumber\\
    &\mathrlap{\cdot\vartheta_{1}(x{+}\tau y|\tau)^{2(N{-}1{-}k{+}n)} \overline{\vartheta_{1}(x{+}\tau y|\tau)}^{2(N{-}1{-}k{+}\bar{n})} \vartheta_{2}(x{+}\tau y|\tau)^{2(k{-}n)} \overline{\vartheta_{2}(x{+}\tau y|\tau)}^{2(k{-}\bar{n})}.}&&
\label{worldsheet binomial expansion}
\end{flalign}

Our objective is now to evaluate the `master' integral
\begin{align}
&{\cal I}_{N_1,\overline{N}_1,N_2,\overline{N}_2}(\tau,\bar\tau) = {\int_0}^{\,1}{dx}{\int_0}^{\,1}{dy}\,e^{{-}4\pi{(}N{-}1{)}\tau_{2}y^{2}}\cdot\nonumber \\
&\cdot\vartheta_{1}{(}x{+}\tau y|\tau{)}^{N_{1}}\overline{\vartheta_{1}{(}x{+}\tau y|\tau{)}}^{\overline{N}_{1}} \vartheta_{2}{(}x{+}\tau y|\tau{)}^{N_{2}}\overline{\vartheta_{2}{(}x{+}\tau y|\tau{)}}^{\overline{N}_{2}},
\label{general worldsheet integral}
\end{align}
with the constraint $N_{1}+N_{2}=2(N{-}1)=\overline{N}_{1}+\overline{N}_{2}$ imposed by level-matching. The integral \eqref{general worldsheet integral} can be computed analytically using the definitions of $\vartheta_{s}$ with $s=1,2$ as in \cite{Kiritsis:2019npv}:
\begin{align}
    \vartheta_s(z|\tau)=\sum_{n\in\mathbb{Z}}q^{\frac{1}{2}\left(n+\frac{1}{2}\right)^2}e^{2\pi i \left(n+\frac{1}{2}\right)\left(z-\frac{r}{2}\right)},\quad q=e^{2\pi i\tau},\quad r=\left\{\begin{aligned}
        &1 \text{ if }s=1\\
        &0 \text{ if }s=2\end{aligned}\right..
    \label{theta definitions worldsheet integral}
\end{align}
The detailed computation is presented in Appendix \ref{appendix worldsheet integral}. Plugging the definitions \eqref{theta definitions worldsheet integral} into the integral \eqref{general worldsheet integral}, we find
\begin{align}
 &{\cal I}_{N_1,\overline{N}_1,N_2,\overline{N}_2}(\tau,\bar\tau) =\prod_{i=1}^{2(N-1)}{\sum_{n_i,\bar{n}_i}}{\int_0}^{\,1}dx{\int_0}^{\,1}dy\,e^{{-}4\pi(N{-}1)\tau_{2}y^{2}}\nonumber \\
 &q^{\frac{1}{2}\left(n_i+\frac{1}{2}\right)^2} e^{2\pi i \left(n_i+\frac{1}{2}\right)\left(x+\tau y-\frac{r_i}{2}\right)}\bar{q}^{\frac{1}{2}\left(\bar{n}_i+\frac{1}{2}\right)^2} e^{-2\pi i \left(\bar{n}_i+\frac{1}{2}\right)\left(x+\bar{\tau} y-\frac{\bar{r}_i}{2}\right)}
\label{general worldsheet integral explicit}
\end{align}
and we see that the integral over $x$ sets $\sum_i n_i = \sum_i \bar{n}_i$. We now choose the parametrization
\begin{align}
    \left\{\begin{aligned}
        &n_i=\ell+K_i\\
        &\bar{n}_i=\ell+\overline{K}_i
    \end{aligned}\right., \quad \ell,\,K_i,\,\overline{K}_i\in\mathbb{Z},
%\label{worldsheet reparametrization}
\end{align}
in such a way that
\begin{align}
    \sum_{i=1}^{2(N-1)} n_i = 2(N-1)\ell + \sum_{i=1}^{2(N-1)} K_i := 2(N-1)\ell + h = \sum_{i=1}^{2(N-1)} \bar{n}_i
\end{align}
with $h:=\sum_i K_i=\sum_i \overline{K}_i\in\{1,\,...,\,2(N-1)\}$. The sum over $\ell$ enables the recasting of the integral over $y$ in \eqref{general worldsheet integral explicit} as a Gaussian integral, so that, in the end, we obtain
\begin{align}
    {\cal I}_{N_{1},\overline{N}_{1},N_{2},\overline{N}_{2}}(\tau,\bar{\tau}) := &\, {(-)^{\frac{N_{1}+\overline{N}_{1}}{2}}\over 2 \sqrt{\tau_2(N{-}1)}}\sum_{h=1}^{2(N-1)}\sum_{\mathbf{K}\in\mathbb{Z}^{2(N-1)}}\sum_{\overline{\mathbf{K}}\in\mathbb{Z}^{2(N-1)}}\delta_{\mathbf{1}_{2(N-1)}\cdot\mathbf{K},h}\,\,\delta_{\mathbf{1}_{2(N-1)}\cdot\overline{\mathbf{K}},h}\nonumber\\
    &\, q^{{1\over 2}\mathbf{K}\cdot\mathbf{K}-{h^{2}\over 4(N-1)}}e^{i\pi\mathbf{a}\cdot\mathbf{K}}\,\,\bar{q}^{{1\over 2}\overline{\mathbf{K}}\cdot\overline{\mathbf{K}}-{h^{2}\over 4(N-1)}}e^{-i\pi\bar{\mathbf{a}}\cdot\overline{\mathbf{K}}},
    \label{integral i}
\end{align}
where
\begin{equation}
\mathbf{K}:=(K_{1},...,K_{2(N-1)})\,,\quad \mathbf{a}:=(\mathbf{1}_{N_{1}},\mathbf{0}_{N_{2}})\,,\quad \mathbf{1}_{\ell}:=\underbrace{(1,...,1)}_{\ell}\,,\quad  \mathbf{0}_{\ell}:=\underbrace{(0,...,0)}_{\ell}.
\end{equation}
We now insert this result into the binomial expansion in eq.~\eqref{worldsheet binomial expansion} and reinstate the modular integral as in eq.~\eqref{self-energy correction}. Factoring $\tau_2^{-{1}/{2}}$ out of $\mathcal{I}_{N_1,\overline{N}_1,N_2,\overline{N}_2}$, we can recombine all the explicit powers of $\tau_2$ appearing so far into $\tau_2^{-5}\cdot\tau_2\cdot\tau_2^{-{1}/{2}}=\tau_2^{-{9}/{2}}$. Hence, we obtain the following closed-form expression for the one-loop modular integral
\begin{align}
    \mathcal{A}^{(N)}=&\, \int_{\cal F}\frac{d^2\tau}{\tau_{2}^{{9}/{2}}\left|\vartheta_1'(0|\tau)\right|^{4(N-1)}} \sum_{k=0}^{N-2} \binom{N-2}{k}^2 \left(\frac{\pi}{\tau_2}\right)^{2(N-2-k)} \sum_{n,\,\bar{n}=0}^{k}\binom{k}{n}\binom{k}{\bar{n}}\nonumber\\
    &\,\tilde{e}_{1}(\tau)^{n}\overline{\tilde{e}_{1}(\tau)}^{\bar{n}} f_{1}(\tau)^{k-n}\overline{f_{1}(\tau)}^{k-\bar{n}}{\cal I}_{2(N-1-k+n),\,2(N-1-k+\bar{n}),\,2(k-n),\,2(k-\bar{n})}(\tau,\bar{\tau}).
    \label{IR divergent integral}
\end{align}
which can be compared with a similar expression in \cite{Chialva:2003hg,Iengo:2002tf,Chialva:2004xm}.

\section{Modular integral: Regularization and  $i\varepsilon$-prescription}
\label{i epsilon section}
The integral \eqref{IR divergent integral} is IR divergent and needs to be regularized for large $\tau_2$ and renormalized. A suitable procedure to achieve the goal has been described by Manschot and Wang in \cite{Manschot:2024prc} and accounts for an extension of the $i\varepsilon$-prescription to string theory \cite{Witten:2013pra, Sen:2016ubf}. Here we briefly outline the results that we will use to regularize our modular integral.\\

The complex structure parameter $\tau$ of the torus takes values in the Teichmüller space $\mathbb{H}\equiv\{\tau_1+i\tau_2\,|\,\tau_1\in\mathbb{R},\,\tau_2>0\}$, namely the upper-half complex plane. The integration domain for $\tau$ is the fundamental region $\mathcal{F}=\mathbb{H}/SL(2,\,\mathbb{Z})$. In particular, we choose  $\mathcal{F}_\infty=\left\{\tau_1+i\tau_2\in\mathbb{H}\,|\,\tau_1\in\left[-\frac{1}{2},\,\frac{1}{2}\right],\,\tau_2\in\left[\sqrt{1-\tau_1^2},\,+\infty\right)\right\}$.\\
For our purposes we need to deal with integrals of the form
\begin{align}
    I_{m,n,s}=\int_\mathcal{F} d^2\tau\, \tau_2^{-s}q^m \bar{q}^n,
    \label{model integral}
\end{align}
where $n,\,m\in\mathbb{R},\,m-n\in\mathbb{Z}$ and $s\in\mathbb{Z}/2$. These integrals are finite for $m+n>0$ or for $m+n=0$ and $s>1$. Instead, the integral diverges for $m+n<0$ when $\tau_2 \to +\infty$, which represents the IR region. Hence, as anticipated, some regularization procedure is required.\\

Firstly, we restrict our attention onto the reduced region \\\begin{equation} \mathcal{F}_{T_0} = \left\{\tau_1\in\left[-\frac{1}{2},\,\frac{1}{2}\right],\, \tau_2\in\left[\sqrt{1-\tau_1^2},\,T_0\right]\right\}\end{equation}
with the understanding that the original integral is recovered by taking the limit $T_0 \to +\infty$. Then, we separate the cutoff-regulated integration domain into two different regions: a small $\mathcal{F}_1$ (with $\tau_2\le 1$) and a large $R_{T_0}=\left\{\tau_1\in\left[-\frac{1}{2},\,\frac{1}{2}\right],\,\tau_2\in\left[1,\,T_0\right]\right\}$. The regulated integral then reads
\begin{align}
    I^{\rm reg}_{m,n,s}(T_0)&=\int_{\mathcal{F}_1} d^2\tau\, \tau_2^{-s}q^m\bar{q}^n + \int_{-\frac{1}{2}}^{\frac{1}{2}} d\tau_1 \int_1^{T_0}d\tau_2\, \tau_2^{-s}q^m\bar{q}^n = \nonumber\\
    &=\int_{-\frac{1}{2}}^{\frac{1}{2}}d\tau_1 \int_{\sqrt{1-\tau_1^2}}^1 d\tau_2\, \tau_2^{-s}q^m\bar{q}^n+\delta_{m,n}[E_s(4\pi m)-T_0^{1-s}E_s(4\pi m T_0)],
    \label{manschot 3.20}
\end{align}
where $E_s(z)$ is the generalized exponential integral
\begin{align}
    E_s(z)=\left\{\begin{aligned}
        &z^{s-1}\int_z^{+\infty}dt\,e^{-t}t^{-s}\quad \text{if }z\in\mathbb{C}\setminus\{0\}\\
        &\frac{1}{s-1}\qquad\qquad\qquad\,\,\,\,\, \text{if }z=0\text{ and }s\neq1\\
        &0\qquad\qquad\qquad\qquad\,\,\,\,\, \text{if }z=0\text{ and }s=1
    \end{aligned}\right..
\end{align}
As a result, the relevant integrals are convergent for $m+n>0$, whereas they generally diverge for $m+n\leq0$. The renormalization consists in removing the third term from eq. \ref{manschot 3.20}, yielding the renormalized integral
\begin{align}
    I^\text{ren}_{m,n,s}=\int_{-\frac{1}{2}}^{\frac{1}{2}} d\tau_1 \int_{\sqrt{1-\tau_1^2}}^1 d\tau_2\, \tau_2^{-s}q^m\bar{q}^n+\delta_{m,n}E_s(4\pi m),\quad m+n\leq0.
\end{align}

If we finally call $F(m,n)$ the coefficient of the integral $I_{m,n,s}$ in an expansion in powers $q^m\bar{q}^n$ (which is exactly the case for our $\mathcal{A}^{(N)}$), we find
\begin{align}
    I^\text{ren}(T_0)=&\sum_{m,n}F(m,n)\int_{-\frac{1}{2}}^{\frac{1}{2}} d\tau_1 \int_{\sqrt{1-\tau_1^2}}^1 d\tau_2\, \tau_2^{-s}q^m\bar{q}^n+\sum_{m\leq0}F(m,m)E_s(4\pi m)+\nonumber\\
    &\quad+\sum_{m>0}F(m,m)[E_s(4\pi m)-T_0^{1-s}E_s(4\pi mT_0)].
    \label{manschot's formula}
\end{align}
This procedure accounts for a consistent analytic continuation. Indeed, when a long tube is developed along the worldsheet, the prescription effectively enables a transition from Euclidean to Lorentzian signature, so that the IR divergences are regulated. The technique was pioneered by Marcus \cite{Marcus:1988vs}, and then revived by Witten \cite{Witten:2013pra}, applied by Sen \cite{Sen:2016ubf,Sen:2016gqt} and Pius \cite{Pius:2013sca,Pius:2014iaa,Pius:2016jsl} to the mass renormalization and then by Stieberger \cite{Stieberger:2023nol}. Extension to the open (super-)string case has been tackled by Mizera and Eberhardt \cite{Eberhardt:2023xck}.\\

In our specific case, the application of the $i\varepsilon$-prescription to eq. \eqref{IR divergent integral} results in 
\begin{equation}
    {\cal A}^{(N)}_{\rm ren} := {\cal A}^{(N)}_{\mathcal{F}_1} + \sum_{\ell=0}^{2(N-2)}\sum_{k_\ell=0}^{\nu_\ell} {c^{(N)}_{\ell,k_\ell}\over \sqrt{N{-}1}} E_{{9\over 2}+\ell}\left(-{\pi b^{(N)}_{k_\ell}\over N{-}1}\right)
    \label{regAmpl}
\end{equation}
where in general we can parametrize the result in terms of the number $\nu_\ell$ of  identical negative powers of $q$ and $\bar{q}$, a rational coefficient $c^{(N)}_{\ell,k_\ell}$ intimately connected to the weight of the various internal channels contributing to the self-energy of the state, a non-negative integer $b_{k_\ell}^{(N)}$ depending on the identical negative powers of $q$ and $\bar{q}$ and finally a finite part which only contributes to the mass shift 
\begin{align}
   {\cal A}^{(N)}_{\mathcal{F}_1}:=&\, \int_{{-}\frac{1}{2}}^{\frac{1}{2}}d\tau_1 \int_{\sqrt{1{-}\tau_1^2}}^1 \frac{d\tau_2}{\tau_{2}^{\frac{9}{2}}\left|\vartheta_1'(0|\tau)\right|^{4(N-1)}} \sum_{k=0}^{N-2} \binom{N{-}2}{k}^2 \left(\frac{\pi}{\tau_2}\right)^{2(N-2-k)} \sum_{n,\,\bar{n}=0}^{k}\binom{k}{n}\binom{k}{\bar{n}}\nonumber\\
    &\,\tilde{e}_{1}(\tau)^{n}\overline{\tilde{e}_{1}(\tau)}^{\bar{n}} f_{1}(\tau)^{k-n}\overline{f_{1}(\tau)}^{k-\bar{n}}{\cal I}_{2(N-1-k+n),\,2(N-1-k+\bar{n}),\,2(k-n),\,2(k-\bar{n})}(\tau,\bar{\tau}).
    \label{smallReg}
\end{align}
The regulated integral over the region $R_{T_0}$ turns out to be suppressed at all mass levels.\\
In the following sections, we will explicitly apply the above procedure to specific cases.

\section{Explicit results}

We are now ready to tackle the mass corrections to FRT states at increasing level $N$. To this aim, we explicitly evaluate the self-energy exploiting the regularization \eqref{regAmpl} for $N=2,3,4$ and then report the results for $N$ up to 10. Notice that massless states at level $N=1$ are stable and are protected against mass corrections. The integral \eqref{smallReg} is evaluated numerically and is observed to converge very quickly to a stable result when truncated up to the second order in both $q$ and $\bar{q}$. This will be explicitly shown for the lightest states in the following.\\
At each mass level, the worldsheet integrals ${\cal I}_{2(N-1-k+n),\,2(N-1-k+\bar{n}),\,2(k-n),\,2(k-\bar{n})}$ were extracted using a Python code. The further integral on the modular parameter was then computed in Wolfram Mathematica.

\subsection{$N=2$ case in Type II}
\label{n=2 section}
The first case where we can find a non-vanishing self-energy is at the first massive level $N=2$. In this particular case, the only integral we have to compute is 
\begin{equation}
    \mathcal{A}^{(2)}= \int_{\cal F}{d^2\tau\over \tau_2^{9\over 2} }\frac{{\cal I}_{2,\,2,\,0,\,0}(\tau,\bar{\tau})}{\left| \vartheta_1'(0|\tau)\right|^{4(N-1)}},
\end{equation}
which reproduces the mass correction and decay width up to an overall constant $C^{(2)}_{\rm II} = (2\pi)^4 g_s^2/(4M_2)$ fixed as in \cite{Marcus:1988vs}.
Following the regularization and renormalization procedure detailed in the previous section, we obtain  
\begin{equation}
 \mathcal{A}^{(2)}_{\rm ren}=A_{\mathcal{F}_1}^{(2)}+\frac{1}{8\pi^4}E_{9\over 2}(0)+\frac{1}{32\pi^4}E_{9\over 2}(-\pi),
\end{equation}
where explicitly we have
\begin{equation}
A_{\mathcal{F}_1}^{(2)}\simeq4\cdot 10^{-4}\,,\quad E_{9\over 2}(0)=0.28571\,,\quad E_{9\over 2}(-\pi)=4.57232 + 14.8433\, i,
\end{equation}
so that 
\begin{equation}
 \mathcal{A}^{(2)}_{\rm ren}=(2.23 + 4.76 \,i)\cdot 10^{-3}
\end{equation}

The imaginary part of the result is exact, since it only depends on the (equal) negative powers of $q$ and $\bar{q}$ (which are always taken into account). Thus, in order to show the efficiency of our truncation, we will look at $\mathrm{Re}\,\mathcal{A}^{(2)}$. Let us label $a_n$ the value of $\mathrm{Re}\,\mathcal{A}^{(2)}$ when truncating the torus integrand up to the order $q^n$, with $n\geq0$. Our measure for the convergence of the mass correction is given by the relative change $r_n:=\left|(a_{n+1}-a_n)/a_n\right|$. We report the first values of $a_n$ and $r_n$ in Table~\ref{N=2 convergence}. Given the small values of $r_n$ already for low $n$, we consider the truncation to the term $q^2\bar{q}^2$ adequate for the accuracy we provide in the present work.
\begin{table}[ht]
\centering
\caption{Values of $a_n:=\mathrm{Re}\,\mathcal{A}^{(2)}$ and the relative change $r_n$ upon truncation of the modular integrand to powers $q^n$ and $\bar{q}^n$.}
\label{N=2 convergence}
%\resizebox{\textwidth}{!}{
\begin{tabular}{|c||c|c|c|c|}
\hline
$n$ & 0 & 1 & 2 & 3\\
\hline
\hline
$a_n\cdot 10^3$ & 2.2390 & 2.2332 & 2.2332 & 2.2332\\
\hline
$r_n$ & - & $2.612903\cdot10^{-3}$ & $2.9334\cdot10^{-5}$ & $2.377149\cdot10^{-7}$\\
\hline
\end{tabular}%}
\end{table}

A few comments are in order. First of all, at level $N=2$, all the states belong in a single super-multiplet consisting of $2^{16}$ states ($2^{15} = (84+44)^2 + 128^2$ bosons in the NS-NS and R-R sectors and  $2^{15} = 2(84+44)\times 128$ fermions in the NS-R and R-NS). Not surprisingly the (one-loop) mass corrections are identical for the components of the supermultiplet. In particular the spin 4 states, truly FRT states, receive the same mass correction as any other state, such as the ones built using the totally antisymmetric tensor ${\cal V}^{(-1)}_C = C_{\mu\nu\rho} \Psi^\mu \Psi^\nu \Psi^\rho e^{-\varphi}$ in the Left- and/or in the Right-moving NS sector \cite{Sen:2016gqt}.\\
Secondly, the sign of the imaginary part is positive in our conventions $S=1+iT$, even though the majority of authors prefer $S=1-iT$  so that ${\rm Im}\delta M <0$, being related to ${\rm Im}T = - TT^\dagger $.\\
Later on we will try and identify a trend of mass corrections and decay widths that implies peculiar stability properties for high-spin FRT states \cite{Chialva:2003hg}. 

\subsection{$N=3$ case in Type II}
\label{n=3 section}
The next case corresponds to $N=3$. The computation is similar to the level $N=2$, but now we have contributions also from the mixed derivatives of the Bargmann kernel. The three binomial sums (with $n$ and $\bar{n}$ ranging from 0 to $k$, and $k$ ranging from 0 to 1) produce four different integrals:
$${\cal I}_{2(N-1-k+n),\,2(N-1-k+\bar{n}),\,2(k-n),\,2(k-\bar{n})}\,\longrightarrow\,\{ {\cal I}_{2,\,2,\,2,\,2};\, {\cal I}_{4,\,2,\,0,\,2};\, {\cal I}_{2,\,4,\,2,\,0};\, {\cal I}_{4,\,4,\,0,\,0}\}.$$
Combining the various contributions and following the renormalization procedure detailed in the Sec. \ref{i epsilon section}, we eventually find
%\begin{align}
% \mathcal{A}^{(3)}_{\rm ren}=&{9\over 64\sqrt{2} \pi^{6}}E_{13\over 2}(0)+{1\over 8 \sqrt{2} \pi^{6}}E_{13\over 2}\left(-{\pi\over 2}\right)+{1\over 256\sqrt{2} \pi^{6}}E_{13\over 2}(-2\pi)\\
%&-{3\over 8 \sqrt{2} \pi^{5}}E_{11\over 2}(0)-{1\over 4 %\sqrt{2} \pi^{5}}E_{11\over 2}\left(-{\pi\over 2}\right)+{1\over 4 \sqrt{2} \pi^{4}}E_{9\over 2}(0)+{1\over 8 \sqrt{2} \pi^{4}}E_{9\over 2}\left(-{\pi\over 2}\right)+\mathcal{A}^{(3)}_{\mathcal{F}_1}
%\end{align}
\begin{align}
    \mathcal{A}^{(3)}_{\rm ren} =& \mathcal{A}^{(3)}_{\mathcal{F}_1} {+} \frac{9}{64\pi^6\sqrt{2}} E_{\frac{13}{2}}(0) {+} \frac{1}{256\pi^6\sqrt{2}} E_{\frac{13}{2}}({-}2 \pi) {+} \frac{1}{8\pi^6\sqrt{2}}E_{\frac{13}{2}}\left({-}\frac{\pi}{2}\right) {+}\nonumber\\
    &{-} \frac{3}{16\pi^5\sqrt{2}}E_{\frac{11}{2}}(0) {-} \frac{1}{8\pi^5\sqrt{2}}E_{\frac{11}{2}}\left({-}\frac{\pi}{2}\right) {+} \frac{1}{8\sqrt{2}\pi^4}E_{\frac{9}{2}}(0) {+} \frac{1}{16\sqrt{2}\pi^4}E_{\frac{9}{2}}\left({-}\frac{\pi}{2}\right),
\end{align}
with the contribution from $\mathcal{F}_1$ given by 
\begin{equation}
    \mathcal{A}^{(3)}_{\mathcal{F}_1}\simeq0.1551\cdot10^{-3}.
\end{equation}
Plugging in the explicit values of the generalized exponential integrals, the final result reads
\begin{equation}
 \mathcal{A}^{(3)}_{\rm ren}=(0.918 + 1.245\, i)\cdot10^{-3},
\end{equation}
up to an overall factor $C^{(3)}_{\rm II} = (2\pi)^4 g_s^2/(4M_3)$.\\

Also in this case, we rely on the real part $a_n:=\mathrm{Re}\,\mathcal{A}^{(3)}$ and the relative change $r_n$ to show the efficiency of our truncation. The first values are listed in Table~\ref{N=3 convergence}. Again, good accuracy is reached already by truncating the modular integrand to the term $q^2\bar{q}^2$.
\begin{table}[ht]
\centering
\caption{Values of $a_n:=\mathrm{Re}\,\mathcal{A}^{(3)}$ and the relative change $r_n$ upon truncation of the modular integrand to powers $q^n$ and $\bar{q}^n$.}
\label{N=3 convergence}
%\resizebox{\textwidth}{!}{
\begin{tabular}{|c||c|c|c|c|}
\hline
$n$ & 0 & 1 & 2 & 3\\
\hline
\hline
$a_n\cdot 10^3$ & 0.9169 & 0.9181 & 0.9181 & 0.9181\\
\hline
$r_n$ & - & $1.2520\cdot10^{-3}$ & $6.7458\cdot10^{-5}$ & $1.0151\cdot10^{-6}$\\
\hline
\end{tabular}%}
\end{table}

Compared to ${N=2}$, we find a decreasing trend in the mass correction. Differently from the previous level, there are two different supermultiplets at level $N=3$. Already in the NS sector (say for the Left-movers) there are many more options, that include
\begin{equation}
    (\partial X)^3,\partial^2 X \partial X, \partial^3 X, \Psi^2  (\partial X)^2, \Psi^2  \partial^2 X, \Psi \partial\Psi \partial X, \partial\Psi \partial\Psi, \Psi \partial^2\Psi, \Psi\Psi\Psi \partial\Psi,\dots,
\end{equation}
with bosonic / fermionic lengths (3,0), (2,0), (1,0); (2,2), (1,2), (1,2)', (0,2), (0,2)', (0,4), that constrain the potential mixing. We plan to investigate these mixings, taking into account Lorentz invariance selection rules and the structure of the supermultiplets.

\subsection{$N=4$ case in Type II}
\label{n=4 section}
The case $N=4$ is more involved. The  two binomial sums (with $n$ and $\bar{n}$ ranging from 0 to 1, 2) produce nine different integrals:
$${\cal I}_{2(N-1-k+n),\,2(N-1-k+\bar{n}),\,2(k-n),\,2(k-\bar{n})}\,\longrightarrow\,$$
$$\{ {\cal I}_{2,\,2,\,4,\,4}; {\cal I}_{4,\,2,\,2,\,4}; {\cal I}_{2,\,4,\,4,\,2}; 
{\cal I}_{4,\,4,\,2,\,2}; {\cal I}_{2,\,6,\,4,\,0}; {\cal I}_{6,\,2,\,0,\,4}; {\cal I}_{6,\,4,\,0,\,2}; 
{\cal I}_{4,\,6,\,2,\,0}; {\cal I}_{6,\,6,\,0,\,0}\}$$
Combining the various contributions and following the renormalization procedure described in the previous section one eventually finds
%$$
% \mathcal{A}^{(4)}_{\rm ren}={25\over 256\sqrt{3} \pi^{8}}E_{17\over 2}(0)+{1\over 4096\sqrt{3} \pi^{8}}E_{17\over 2}(-3\pi)+{9\over 512 \sqrt{3} \pi^{8}}E_{17\over 2}\left(-{4\pi\over 3}\right)
%$$
%$$
%+{255\over 2048 \sqrt{3} \pi^{8}}E_{17\over 2}\left(-{\pi\over 3}\right)-{15\over 32\sqrt{3} \pi^{7}}E_{15\over 2}(0)-{3\over 64 \sqrt{3} \pi^{7}}E_{15\over 2}\left(-{4\pi\over 3}\right)-{15\over 32 \sqrt{3} \pi^{7}}E_{15\over 2}\left(-{\pi\over 3}\right)
%$$
%$$
%+{7\over 8\sqrt{3} \pi^{6}}E_{13\over 2}(0)+{1\over 32 \sqrt{3} \pi^{6}}E_{13\over 2}\left(-{4\pi\over 3}\right)+{47\over 64 \sqrt{3} \pi^{6}}E_{13\over 2}\left(-{\pi\over 3}\right)-{3\over 4\sqrt{3} \pi^{5}}E_{11\over 2}(0)
%$$
%\begin{equation}
%-{1\over 2 \sqrt{3} \pi^{5}}E_{11\over 2}\left(-{\pi\over 3}\right)+{1\over 4\sqrt{3} \pi^{4}}E_{9\over 2}(0)+{1\over 8 \sqrt{3} \pi^{4}}E_{9\over 2}\left(-{\pi\over 3}\right)+{\cal A}^{(4)}_{\mathcal{F}_1}
%\end{equation}
\begin{align}
    \mathcal{A}^{(4)}_{\rm ren} =& \mathcal{A}^{(4)}_{\mathcal{F}_1} {+} \frac{25\sqrt{3}}{256\pi^8}E_{\frac{17}{2}}(0) {+} \frac{\sqrt{3}}{4096\pi^8}E_{\frac{17}{2}}(-3\pi) {+} \frac{9\sqrt{3}}{512\pi^8}E_{\frac{17}{2}}\left({-}\frac{4\pi}{3}\right) {+}\nonumber\\
    &{+} \frac{225\sqrt{3}}{2048\pi^8}E_{\frac{17}{2}}\left(-\frac{\pi}{3}\right) {-} \frac{15\sqrt{3}}{64\pi^7}E_{\frac{15}{2}}(0) {-} \frac{3\sqrt{3}}{128\pi^7}E_{\frac{15}{2}}\left({-}\frac{4\pi}{3}\right) {-} \frac{15\sqrt{3}}{64\pi^7}E_{\frac{15}{2}}\left({-}\frac{\pi}{3}\right) {+}\nonumber\\
    &{+} \frac{23}{32\sqrt{3}\pi^6}E_{\frac{13}{2}}(0) {+} \frac{1}{32\sqrt{3}\pi^6}E_{\frac{13}{2}}\left({-}\frac{4\pi}{3}\right) {+} \frac{79}{128\sqrt{3}\pi^6}E_{\frac{13}{2}}\left({-}\frac{\pi}{3}\right) {-} \frac{\sqrt{3}}{8\pi^5}E_{\frac{11}{2}}(0) {+}\nonumber\\
    &{-} \frac{1}{4\sqrt{3}\pi^5}E_{\frac{11}{2}}\left({-}\frac{\pi}{3}\right) {+} \frac{1}{8\sqrt{3}\pi^4}E_{\frac{9}{2}}(0) {+} \frac{1}{16\sqrt{3}\pi^4}E_{\frac{9}{2}}\left({-}\frac{\pi}{3}\right).
\end{align}
where the integral over $\mathcal{F}_1$ produces
\begin{equation}
{\cal A}^{(4)}_{\mathcal{F}_1}\simeq 0.0907 \cdot 10^{-3}
\end{equation}
and, plugging in the exact values of the generalized exponential integrals, we obtain
\begin{equation}
 \mathcal{A}^{(4)}_{\rm ren}=(0.487 + 0.614\,i)\cdot 10^{-3},
\end{equation}
up to an overall factor $C^{(4)}_{\rm II} =(2\pi)^4 g_s^2 /(4M_4)$.\\

The efficiency of the truncation is proven analogously to the previous cases, using the relative change $r_n$ of $a_n:=\mathrm{Re}\,\mathcal{A}^{(4)}$. The results are summarized in Table~\ref{N=4 convergence}.
\begin{table}[ht]
\centering
\caption{Values of $a_n:=\mathrm{Re}\,\mathcal{A}^{(4)}$ and the relative change $r_n$ upon truncation of the modular integrand to powers $q^n$ and $\bar{q}^n$.}
\label{N=4 convergence}
%\resizebox{\textwidth}{!}{
\begin{tabular}{|c||c|c|c|c|}
\hline
$n$ & 0 & 1 & 2 & 3\\
\hline
\hline
$a_n\cdot 10^3$ & 0.4854 & 0.4871 & 0.4871 & 0.4871\\
\hline
$r_n$ & - & $3.4438\cdot10^{-3}$ & $1.0394\cdot10^{-4}$ & $2.4734\cdot10^{-6}$\\
\hline
\end{tabular}%}
\end{table}

Compared to $N=2$ and $N=3$, we find confirmation of a decreasing trend in the mass correction. In this case, already for purely bosonic vertex operators one could study the mixing between $\partial^2 X \partial^2 X$ and $\partial^3 X\partial X$, allowed by the `conservation' of the bosonic and fermionic lengths in the Left- (and similarly in the Right-) moving sector. 

\subsection{Summary and plot for FRT states at level $N=5,\dots,10$}
\label{summary of results}
\begin{figure}[h!]
    \centering
    \includegraphics[width=0.9\linewidth]{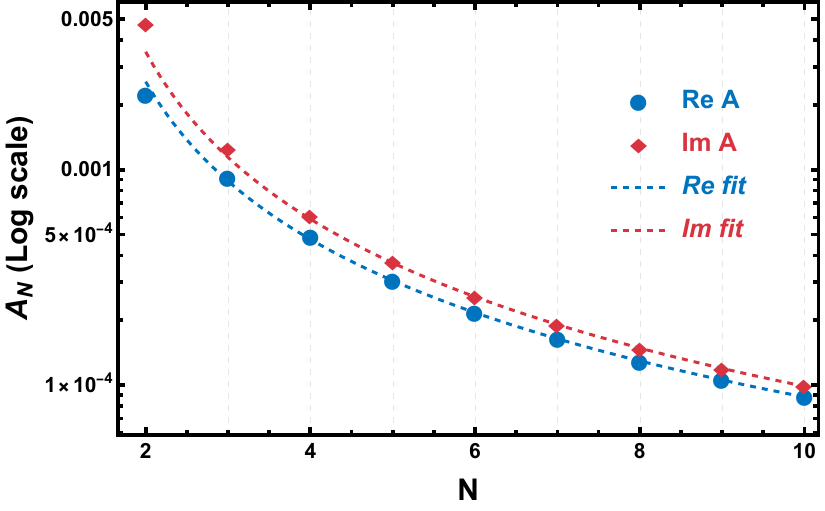}
    \caption{Fit of the real and imaginary part of $\mathcal{A}^{(N)}$ with the mass level.}
    \label{fit figure}
\end{figure}

As the level increases, we find more and more integrals to compute, in fact $(N-1)^2$ or rather $N(N-1)/2$ up to the exchange of the holomorphic and anti-holomorphic parts, {\it i.e.} ($N_1$, $N_2$) $\leftrightarrow$ ($\overline{N}_1$, $\overline{N}_2$), that keeps $N_1+N_2= 2(N-1) =\overline{N}_1+\overline{N}_2$ fixed. Nevertheless one can carry out the calculations without excessive effort up to level $N=10$ (with 45 independent integrals, rather than 81). We refrain from displaying the individual results, that we rather collect in Table~\ref{tab:reimA} and represent in Figure~\ref{fit figure}. Our results are in agreement with those obtained by Baccianti, Eberhardt and Mizera in \cite{Baccianti:2025whd}, where they consider the one-loop 4-graviton amplitude through an alternative method to evaluate modular integrals \cite{Baccianti:2025gll}.

\begin{table}[ht]
\centering
\caption{Values of $\mathrm{Re}\,\mathcal{A}^{(N)}$ and $\mathrm{Im}\,\mathcal{A}^{(N)}$ in units of $10^{-3}C^{(N)}$ for $N=2,\dots,10$.}
\label{tab:reimA}
\resizebox{\textwidth}{!}{
\begin{tabular}{|c||c|c|c|c|c|c|c|c|c|}
\hline
$N$ & 2 & 3 & 4 & 5 & 6 & 7 & 8 & 9 & 10 \\
\hline
\hline
$\mathrm{Re}\,{\cal A}$ & 2.2332 & 0.9181 & 0.4871 & 0.3065 & 0.2172 & 0.1642 & 0.1292 & 0.1058 & 0.0885\\
\hline
$\mathrm{Im}\,{\cal A}$ & 4.7619 & 1.2446 & 0.6143 & 0.3759 & 0.2571 & 0.1914 & 0.1484 & 0.1197 & 0.0991\\
\hline
\end{tabular}}
\end{table}
It is tempting to extrapolate the result to large $N$ and fit the data with a power law, yielding
\begin{equation}
    \mathrm{Re}\mathcal{A}^{(N)}\approx \frac{A_1}{(N-1)^{\beta_1}}\quad\text{and}\quad \mathrm{Im}\mathcal{A}^{(N)}\approx \frac{A_2}{(N-1)^{\beta_2}},
\end{equation}
where $A_1=2.56\cdot10^{-3}$, $\beta_1=1.53\approx 3/2$, $A_2=3.52\cdot10^{-3}$ and $\beta_2=1.63\approx8/5$.\\

We can also use these results to infer a scaling relation for the mass correction $\mathrm{Re}\delta M$ and the decay width $\mathrm{Im}\delta M:=\Gamma$, provided that we multiply times the factor 
\begin{equation}
C^{(N)}_{\rm II} = \frac{(2\pi)^4g_s^2}{4M_N},
\end{equation}
with $\alpha' M_N^2 = 4(N{-}1)$ and $g_s$ related to the 10-dimensional Newton constant by
\begin{equation}
\kappa^2_{10} = 8\pi G_N^{(10)} = {g_s^2 {\alpha'}^4\over (2\pi)^{7}}.
\end{equation}
Both the exponents $\beta_1$ and $\beta_2$ would gain an extra contribution, thus yielding the scaling relations $\mathrm{Re}\delta M\sim N^{-2.03}\approx N^{-2}$ and $\Gamma\sim N^{-2.13}$. Once again, notice the decreasing trend in both the mass corrections and decay widths. Compared to previous estimates based on the breaking probability of long strings \cite{Chialva:2003hg,Iengo:2002tf,Damour:1999aw}, we observe a faster decrease with $N$, with our limited data. We remark that the scaling laws provided above have been obtained by considering only the first ten mass levels and should thus be regarded as an intermediate, finite-$N$ result, rather than an asymptotic statement. We postpone a more thorough analysis of higher-mass states to future works.

\section{Conclusions and outlook}
\label{conclusions}

We have approached a long-term systematic investigation of one-loop mass corrections and mixing in string theory, based on the direct computation of self-energy corrections. In order to avoid issues with vacuum stability (bosonic string tachyon) and/or subtle phases that are remnant of the projected-out tachyon (heterotic string), we started with FRT states with maximal spin in the NS-NS sector of the Type II superstring, that cannot mix with any other states in perturbation theory, due to Lorentz symmetry (and SUSY). Physical vertex operators for these states are easy to write down both in  the covariant and the DDF approach, that we adapted to our present needs. The worldsheet integrals were performed analytically, exploiting remarkable properties of elliptic functions and character decomposition. The final modular integrals over the fundamental domain turned out to be IR divergent in their real part, related to mass corrections, and were performed numerically using the method developed by Manschot and Wang \cite{Manschot:2024prc} and based on the $i\varepsilon$ prescription in string theory revived by Witten \cite{Witten:2013pra} and then by Sen, Pius, Eberhardt and Mizera for specific computations (similar to the ones here) \cite{Sen:2016ubf,Sen:2016gqt,Pius:2014iaa,Pius:2016jsl,Pius:2013sca,Eberhardt:2023xck}. So far we have only addressed FRT states up to level $N=10$ and we found a clear trend of decreasing mass corrections, related to ${\rm Re}{\cal A}^{(N)}_{FRT}$ and decay widths, related to ${\rm Im}{\cal A}^{(N)}_{FRT}$. Our results agree with those obtained in \cite{Baccianti:2025whd} through a different approach, thus strengthening the consistency of both methods.\\

We plan to continue our investigation and study mass corrections and mixing among more general massive Type II superstring states in the NS-NS sector. Thanks to SUSY, this should be sufficient to identify the structure of the one-loop mass matrix at arbitrary level \cite{Hanany:2010da,Bianchi:2003wx,Beisert:2003te}. We hope to be able soon to give support to our `conjecture' that the (type II) superstring spectrum be governed by some random matrix theory, in spite of the tight constraints imposed by the integrable nature of the worldsheet dynamics. This should not come as a surprise, since the chaotic nature of the spectrum of anomalous dimensions of N=4 SYM that is dual to Type IIB on $AdS_5{\times}S^5$ has been argued to hold \cite{McLoughlin:2022jyt}, notwithstanding the integrability of the `super-spin-chain' underlying the tree-level (sphere) mixing of operators in the planar limit. An early exploration of the asymptotic regime of one-loop string amplitudes was carried out in \cite{Baccianti:2026lpc}, proving the current interest of the community in the topic.\\
As we have mentioned, FRT states do not exhibit perturbative mixing, as a consequence of their Lorentz representation. Thus, they do not point themselves at a random matrix theory description of mass corrections. Yet, we consider them a consistent starting point for a systematic study of such effects, in which we have been able to develop the operational toolkit that we intend to use in the future.\\
Further circumstantial support to our conjecture comes from recent investigations of the dynamics of highly excited string states that have revealed interesting chaotic properties \cite{Gross:2021gsj,Bianchi:2022mhs,Firrotta:2023wem,Hashimoto:2022bll,Bianchi:2023uby}.
That an intricate mixing should take place among generic string states, leading to a chaotic spectrum with level repulsion, was originally motivated by the quest for a string description of hadronic resonances. The quest for a string description of black-hole micro-states provides additional strong motivation in this endeavor. 

\acknowledgments
We would like to thank F. Alday, C. Bachas, M. M. Baccianti, S. Cacciatori, P. Di Vecchia, L. Eberhardt, E. Kiritsis, J. Manschot, C. Markou, S. Mizera, F. Morales, V. Niarchos, I. Pesando, R. Russo, J. Sonnenschein, J. Troost, G. Veneziano and D. Weissman for useful discussions and comments.\\
The work of MB is partially supported by the INFN network ST\&FI.\\
MF wants to thank CERN for their hospitality during a period when part of the research work has been conducted.\\
LG would like to thank the École Normale Supérieure for their hospitality during the final stages of this research work. He is grateful to M. Dirindin for helpful advice on the optimization of our codes.

\newpage
\begin{appendix}

\section{Useful definitions and series expansions}
\label{appendix formulae}
{\bf Jacobi $\boldsymbol{\vartheta}$ functions}\\
Our definitions of the Jacobi $\vartheta$ functions follow \cite{Kiritsis:2019npv}. In particular, the $\theta\!\left[\begin{smallmatrix} a \\ b \end{smallmatrix}\right](z|\tau)$ function is defined as
\begin{align}
    \theta\!\left[\begin{smallmatrix} a \\ b \end{smallmatrix}\right](z|\tau) := \sum_{n\in\mathbb{Z}} q^{\frac{1}{2}\left(n-\frac{a}{2}\right)^2}e^{2\pi i\left(z-\frac{b}{2}\right)\left(n-\frac{a}{2}\right)},\qquad q:=e^{2\pi i\tau}.
    \label{theta definition appendix}
\end{align}
In the usual notation, we have
\begin{align}
    \vartheta_1(z|\tau) := \theta\!\left[\begin{smallmatrix} 1 \\ 1 \end{smallmatrix}\right](z|\tau) &= \sum_{n\in\mathbb{Z}} q^{\frac{1}{2}\left(n-\frac{1}{2}\right)^2}e^{2\pi i\left(z-\frac{1}{2}\right)\left(n-\frac{1}{2}\right)} = \nonumber\\
    &=2q^{\frac{1}{8}}\sin(\pi z)\prod_{n=1}^{+\infty}(1-q^n)\left(1-q^ne^{2\pi iz}\right)\left(1-q^ne^{-2\pi iz}\right),
    \label{theta 1 appendix}
\end{align}
\begin{align}
    \vartheta_2(z|\tau) := \theta\!\left[\begin{smallmatrix} 1 \\ 0 \end{smallmatrix}\right](z|\tau) &= \sum_{n\in\mathbb{Z}} q^{\frac{1}{2}\left(n-\frac{1}{2}\right)^2}e^{2\pi iz\left(n-\frac{1}{2}\right)} = \nonumber\\
    &=2q^{\frac{1}{8}}\cos(\pi z)\prod_{n=1}^{+\infty}(1-q^n)\left(1+q^ne^{2\pi iz}\right)\left(1+q^ne^{-2\pi iz}\right),
    \label{theta 2 appendix}
\end{align}
\begin{align}
    \vartheta_3(z|\tau) := \theta\!\left[\begin{smallmatrix} 0 \\ 0 \end{smallmatrix}\right](z|\tau) &= \sum_{n\in\mathbb{Z}} q^{\frac{n^2}{2}}e^{2\pi izn} = \nonumber\\
    &=\prod_{n=1}^{+\infty}(1-q^n)\left(1+q^{n-\frac{1}{2}}e^{2\pi iz}\right)\left(1+q^{n-\frac{1}{2}}e^{-2\pi iz}\right),
    \label{theta 3 appendix}
\end{align}
\begin{align}
    \vartheta_4(z|\tau) := \theta\!\left[\begin{smallmatrix} 0 \\ 1 \end{smallmatrix}\right](z|\tau) &= \sum_{n\in\mathbb{Z}} q^{\frac{n^2}{2}}e^{2\pi i\left(z-\frac{1}{2}\right)n} = \nonumber\\
    &=\prod_{n=1}^{+\infty}(1-q^n)\left(1-q^{n-\frac{1}{2}}e^{2\pi iz}\right)\left(1-q^{n-\frac{1}{2}}e^{-2\pi iz}\right)
    \label{theta 4 appendix}
\end{align}
At $z=0$, we use the series expansions
\begin{align}
    \vartheta_2(0|\tau) = 2q^{\frac{1}{8}}\left[1 + q + q^3 + q^6 + q^{10} + q^{15} + \mathcal{O}\left(q^{20}\right)\right],
    \label{theta 2 expansion}
\end{align}
\begin{align}
    \vartheta_3(0|\tau) = 1 + 2q^{\frac{1}{2}} + 2q^2 + 2q^{\frac{9}{2}} + 2q^8 + 2q^{\frac{25}{2}} + 2 q^{18} + \mathcal{O}\left(q^{\frac{41}{2}}\right),
    \label{theta 3 expansion}
\end{align}
\begin{align}
    \vartheta_4(0|\tau) = 1 - 2q^{\frac{1}{2}} + 2q^2 - 2q^{\frac{9}{2}} + 2q^8 - 2q^{\frac{25}{2}} + 2 q^{18} + \mathcal{O}\left(q^{\frac{41}{2}}\right).
    \label{theta 4 expansion}
\end{align}
The Dedekind $\eta(\tau)$ function is
\begin{align}
    \eta(\tau):=q^{\frac{1}{24}}\prod_{n=1}^{+\infty}(1-q^n) = q^{\frac{1}{24}} \left[1 - q - q^2 + q^5 + q^7 - q^{12} - q^{15} + \mathcal{O}\left(q^{20}\right)\right]
    \label{eta definition appendix}
\end{align}
and it is related to $\partial_z\vartheta_1(z|\tau):=\vartheta_1'(z|\tau)$ through
\begin{align}
    \vartheta_1'(0|\tau)=2\pi\eta(\tau)^3 = 2\pi q^{\frac{1}{8}} \left[1 - 3q + 5q^3 - 
 7q^6 + 9q^{10} - 
 11q^{15} + \mathcal{O}\left(q^{20}\right)\right].
    \label{theta1 and eta appendix}
\end{align}
{\bf Type II partition function}\\
The Type II partition function is computed as
\begin{align}
{\cal Z}(\tau,\bar{\tau}) &= \int \frac{d^8p}{(2\pi)^8}e^{-\pi\al\tau_2 p^2}\text{Tr}\left[\frac{1-(-)^{F_L}}{2}\frac{1-(-)^{F_R}}{2}q^{\hat{L}_0}\bar{q}^{\bar{\hat{L}}_0}\right],
\end{align}
where the trace is computed over the full Hilbert space and $F_L$ and $F_R$ count the number of left- and right-moving fermionic excitations, respectively. The momentum integral gives
\begin{align}
    \int \frac{d^8p}{(2\pi)^8}e^{-\pi\al\tau_2 p^2} = \left(\frac{1}{4\pi^2\al\tau_2}\right)^4,
\end{align}
whereas the trace reconstructs Jacobi's abstruse identity for $\vartheta$ functions. For our purposes, it is useful to factor out overall constants and redefine a spin-structure-dependent term
\begin{align}
    \mathcal{Z}_{(s_1,s_2)}(\tau,\bar{\tau}):=\frac{\vartheta_{s_1}(0|\tau)^4 \overline{\vartheta_{s_2}(0|\tau)}^4}{4\tau_2^4 \eta(\tau)^{12}\overline{\eta(\tau)}^{12}},
\end{align}
which now undergoes a summation over $(s_1,\,s_2)$.\\
\newline
{\bf Bosonic and fermionic propagators}\\
On a torus with modular parameter $\tau=\tau_1+i\tau_2$, the worldsheet coordinate can be parametrized as $z=x+\tau y=x+\tau_1 y + i\tau_2 y$, with $(x,\,y)\in[0,\,1]\times[0,\,1]$.\\
The bosonic propagator (Bargmann kernel) arising from the contraction of two bosonic fields $X(z,\bar{z})$ reads
\begin{align}
\langle X(z,\bar{z})X(w,\bar{w}) \rangle_{T^2}&=: {G}(z-w,\bar{z}-\bar{w}|\tau)= \nonumber\\
&=-\log \Big| {\vartheta_1(z-w|\tau) \over \vartheta'_1(0|\tau)}   \Big|^{2} + 2\pi{[\mathrm{Im} (z-w)]^2 \over  \mathrm{Im}\,\tau}.
\end{align}
The fermionic propagator (Szego kernel) is instead
\begin{equation}
\begin{split}
\langle \Psi(z) \Psi(w) \rangle_{{\cal T}}^{(s)}=\begin{cases} S_s(z-w)= \frac{\vartheta_s(z-w|\tau)}{\vartheta_1(z-w|\tau)}  {\vartheta_1'(0|\tau) \over \vartheta_s(0|\tau)} \,, \quad s=2,3,4  \\
S(z-w)=  -\partial_{z}\langle X(z)X(w) \rangle_{T^2} \,, \,\,\,\, s=1\end{cases}
\end{split}.
\end{equation}
\newline
{\bf Weierstrass elliptic $\boldsymbol{\wp}$ function}\\
On the torus lattice, the Weierstrass elliptic $\wp(z|\tau)$ function can be written as
\begin{align}
    \wp(z|\tau) = \frac{1}{z^{2}} + \sum_{(n,m)\neq 0}\left[\frac{1}{(z + n + m\tau)^{2}} - \frac{1}{(n + m\tau)^{2}}\right]
    \label{wp definition appendix}
\end{align}
and satisfies the differential equation
\begin{align}
    [\wp'(z|\tau)]^2 = 4\wp(z|\tau)^3-2\left(e_1^2 +e_2^2+e_3^2\right)\wp(z|\tau)-4e_1e_2e_3,
\end{align}
where the lattice roots $e_i$ are
\begin{align}
    e_i:=-4\pi i\partial_\tau\log\frac{\vartheta_{i+1}(0|\tau)}{\eta(\tau)},\quad i=1,\,2,\,3.
    \label{lattice roots appendix}
\end{align}
Introducing the modular form
\begin{align}
    \mathcal{E}(\tau)=4\pi i\partial_\tau\log\left[\eta(\tau)\sqrt{\tau_2}\right],
    \label{modular form E appendix}
\end{align}
the Bargmann kernel can be rewritten as
\begin{align}
    \partial_z^2 G(z|\tau)=\wp(z|\tau)-\mathcal{E}(\tau)
\end{align}
and, since
\begin{align}
    \wp(z|\tau) = e_i(\tau)+S_{i+1}^2(z|\tau) &= e_i(\tau)+\frac{\vartheta_{i+1}(z|\tau)^2\left[\vartheta_1'(0|\tau)\right]^2}{\vartheta_1(z|\tau)^2\vartheta_{i+1}(0|\tau)^2} :=\nonumber\\
    &=e_i(\tau)+f_i(\tau)\frac{\vartheta_{i+1}(z|\tau)^2}{\vartheta_1(z|\tau)^2},
\end{align}
we find
\begin{align}
    \partial_z^2 G(z|\tau) = \tilde{e}_i(\tau) + f_i(\tau)\frac{\vartheta_{i+1}(z|\tau)^2}{\vartheta_1(z|\tau)^2},
\end{align}
where we have defined
\begin{align}
    \tilde{e}_i(\tau) := e_i(\tau)-\mathcal{E}(\tau).
\end{align}
In particular, we often refer to
\begin{align}
    \tilde{e}_1(\tau) &= -4\pi i\partial_\tau\log\left[\vartheta_2(0|\tau)\sqrt{\tau_2}\right] = \nonumber\\
    &=\pi\left(\pi - \frac{1}{\tau_2}\right) + 8\pi^2 q\left[1 - q + 4q^2 - 
 5q^3 + 6q^4 - 4q^5 + \mathcal{O}(q^6)\right]
    \label{tilde e 1 appendix}
\end{align}
and
\begin{align}
    f_1(\tau) = \frac{\vartheta_1'(0|\tau)^2}{\vartheta_2(0|\tau)^2} = 8\pi^2 \left[\frac{1}{8} - q + 3q^2 - 4q^3 + 
 3q^4 - 6q^5 + 12q^6 + 
 \mathcal{O}(q^7)\right].
\end{align}

\section{Full analytic computation of the worldsheet integral}
\label{appendix worldsheet integral}
The worldsheet contribution to the amplitude comes from the general integral \eqref{worldsheet integral}
\begin{flalign}
    \mathrlap{{\int_0}^{\,1}{dx}{\int_0}^{\,1}{dy}\,e^{{-}4\pi{(}N{-}1{)}\tau_{2}y^{2}} \vartheta_{1}{(}x{+}\tau y|\tau{)}^{N_{1}}\overline{\vartheta_{1}{(}x{+}\tau y|\tau{)}}^{\overline{N}_{1}} \vartheta_{2}{(}x{+}\tau y|\tau{)}^{N_{2}}\overline{\vartheta_{2}{(}x{+}\tau y|\tau{)}}^{\overline{N}_{2}},}&&
    \label{general worldsheet integral appendix}
\end{flalign}
with the level-matching constraint $N_1+N_2 = 2(N-1) = \overline{N}_1+\overline{N}_2$. As we stated in Section \ref{worlsheet integration section}, we can compute this integral analytically by means of the definitions
\begin{align}
    \vartheta_s(z|\tau)=\sum_{n\in\mathbb{Z}}q^{\frac{1}{2}\left(n+\frac{1}{2}\right)^2}e^{2\pi i \left(n+\frac{1}{2}\right)\left(z-\frac{r}{2}\right)},\quad q=e^{2\pi i\tau},\quad r=\left\{\begin{aligned}
        &1 \text{ if }s=1\\
        &0 \text{ if }s=2\end{aligned}\right.,
    \label{theta definitions worldsheet integral appendix}
\end{align}
turning the integral \eqref{general worldsheet integral appendix} into
\begin{align}
    \prod_{i=1}^{2(N-1)}{\sum_{n_i,\bar{n}_i}}{\int_0}^{\,1}dx{\int_0}^{\,1}dy\,e^{{-}4\pi(N{-}1)\tau_{2}y^{2}} &\,q^{\frac{1}{2}\left(n_i+\frac{1}{2}\right)^2} e^{2\pi i \left(n_i+\frac{1}{2}\right)\left(x+\tau y-\frac{r_i}{2}\right)}\nonumber\\
    &\,\bar{q}^{\frac{1}{2}\left(\bar{n}_i+\frac{1}{2}\right)^2} e^{-2\pi i \left(\bar{n}_i+\frac{1}{2}\right)\left(x+\bar{\tau} y-\frac{\bar{r}_i}{2}\right)}
\label{general worldsheet integral explicit appendix}
\end{align}
Since all the $n_i$s are integers, the integral over $x$ results in the condition
\begin{equation}
    \sum_{i=1}^{2(N-1)} n_i = \sum_{i=1}^{2(N-1)} \bar{n}_i
    \label{delta constraint n_i}
\end{equation}

The integral \eqref{general worldsheet integral explicit appendix} ultimately depends on the $n_i$s through their sum. Given that there are $2(N-1)$ such indices, each spanning over $\mathbb{Z}$, our aim is to reparametrize this sum through a multiple of $2(N-1)$ and an additional contribution spanning over $\mathbb{Z}_{2(N-1)}$. The correct way to do this is to take into account the degeneracy of each value of $\sum_i n_i$. To this end, we define the parametrization
\begin{align}
    \left\{\begin{aligned}
        &n_i=\ell+K_i\\
        &\bar{n}_i=\ell+\overline{K}_i
    \end{aligned}\right., \quad \ell,\,K_i,\,\overline{K}_i\in\mathbb{Z},
    \label{worldsheet reparametrization appendix}
\end{align}
which naturally leads to
\begin{align}
    \sum_i n_i = 2(N-1)\ell + \sum_i K_i\equiv 2(N-1)\ell + h = 2(N-1)\ell + \sum_i \overline{K}_i = \sum_i \bar{n}_i
\end{align}
with $h:=\sum_i K_i=\sum_i \overline{K}_i\in\{1,...,2(N-1)\}$. Notice that, while the sum over $h$ is now defined on $\mathbb{Z}_{2(N-1)}$, the degeneracy mentioned above is still captured by the sums over the $K_i$s, which are defined on $\mathbb{Z}$. The sums appearing in \eqref{general worldsheet integral explicit appendix} are thus turned into
\begin{align}
    \prod_{i=1}^{2(N-1)}\sum_{n_i\in\mathbb{Z}}\sum_{\bar{n}_i\in\mathbb{Z}} \quad\longrightarrow\quad \sum_{\ell\in\mathbb{Z}}\sum_{h=1}^{2(N-1)}\prod_{i=1}^{2(N-1)}\sum_{K_i\in\mathbb{Z}} \delta_{\sum_i K_i,\,h} \sum_{\overline{K}_i\in\mathbb{Z}} \delta_{\sum_i \overline{K}_i,\,h}.
\end{align}

The full exponent of the integrand in \eqref{general worldsheet integral explicit appendix} reads
\begin{align}
    &-4(N-1)\pi\tau_2 y^2 + 2\pi i\sum_i \left(n_i+\frac{1}{2}\right)(i\tau_2 y) - 2\pi i\sum_i\left(\bar{n}_i+\frac{1}{2}\right)(-i\tau_2 y) + \nonumber\\
    & + \pi i\tau\sum_i\left(n_i+\frac{1}{2}\right)^2 - \pi i\bar{\tau}\sum_i\left(\bar{n}_i+\frac{1}{2}\right)^2 + \nonumber\\
    & + 2\pi i\sum_i\left(n_i+\frac{1}{2}\right)\left(-\frac{r_i}{2}\right) - 2\pi i\sum_i\left(\bar{n}_i+\frac{1}{2}\right)\left(-\frac{\bar{r}_i}{2}\right).
\end{align}
By definition, we have $\sum_i r_i = N_1$ and $\sum_i \bar{r}_i = \overline{N}_1$. If we plug in the reparametrization \eqref{worldsheet reparametrization appendix}, we obtain
\begin{align}
    & -4(N-1)\pi\tau_2 y^2 - 4\pi\tau_2 y\left[2(N-1)\left(\ell+\frac{1}{2}\right)+\sum_i K_i\right] + \nonumber\\
    & + i\pi\tau_1\left[\sum_i\left(\ell+K_i+\frac{1}{2}\right)^2-\sum_i\left(\ell+\bar{K}_i+\frac{1}{2}\right)^2 \right] +\nonumber\\
    & - \pi\tau_2\left[\sum_i\left(\ell+K_i+\frac{1}{2}\right)^2+\sum_i\left(\ell+\bar{K}_i+\frac{1}{2}\right)^2 \right] +\nonumber\\
    & - i\pi\sum_i(\ell+K_i)r_i + i\pi\sum_i(\ell+\bar{K}_i)\bar{r}_i
\end{align}
We then complete the square with the term $-4(N-1)\pi\tau_2\left(\ell+\frac{1}{2}+\frac{\sum_i K_i}{2(N-1)}\right)^2$, so that the exponent ultimately reads
\begin{align}
    &-4(N-1)\pi\tau_2\left(y+\ell+\frac{1}{2}+\frac{\sum_i K_i}{2(N-1)}\right)^2 -\pi\tau_2\left(\sum_i K_i^2 + \sum_i \overline{K}_i^2\right) + \nonumber\\
    & + \pi\tau_2\frac{\left(\sum_i K_i\right)^2}{N-1} + i\pi\tau_1 \left(\sum_i K_i^2 - \sum_i \overline{K}_i^2\right) - i\pi\left(\ell+\frac{1}{2}\right)(N_1-\overline{N}_1) + \nonumber\\
    & - i\pi\sum_i K_ir_i +i\pi\sum_i \overline{K}_i\bar{r}_i.
    \label{final exponent appendix}
\end{align}
The index $\ell$ only appears within the first term. Therefore, we can change the integration variable from $y$ to $y_\ell := y+\ell$ and sum over $\ell$ afterward, so that the integral over $y$ becomes a Gaussian integral:
\begin{align}
    &\,\sum_{\ell\in\mathbb{Z}}\int_0^1 dy \, \exp\left[-4\pi\tau_2(N-1)\left(y+\ell+\frac{1}{2}+\frac{\sum_i K_i}{2(N-1)}\right)^2\right] = \nonumber\\
    =&\,\sum_{\ell\in\mathbb{Z}}\int_\ell^{1+\ell} dy_\ell \, \exp\left[-4\pi\tau_2(N-1)\left(y_\ell+\frac{1}{2}+\frac{\sum_i K_i}{2(N-1)}\right)^2\right] = \nonumber\\
    =&\,\int_{\mathbb{R}}dy_\ell \, \exp\left[-4\pi\tau_2(N-1)\left(y_\ell+\frac{1}{2}+\frac{\sum_i K_i}{2(N-1)}\right)^2\right] = \frac{1}{2\sqrt{\tau_2 (N-1)}}.
\end{align}
Finally, we observe that the $r_i$s appearing in the exponent \eqref{final exponent appendix} effectively select $N_1$ of the $K_i$s (and analogously for the $\bar{r}_i$s), yielding an overall sign for the specific combination of the $K_i$s. For this reason, it is convenient to make a distinction among the $K_i$s: we label $k_i$ the indices related to the powers of $\vartheta_1$, $\hat{k}_i$ those of the powers of $\vartheta_2$, and analogously for the anti-holomorphic sector. By further reinstating the usual variables $q=e^{2\pi i\tau}$ and $\bar{q}$, the integral takes the form
\begin{align}
    {\cal I}_{N_{1},\overline{N}_{1},N_{2},\overline{N}_{2}}(\tau,\bar{\tau}) {:=}& {(-)^{\frac{N_{1}}{2}+\frac{\overline{N}_{1}}{2}}\over 2 \sqrt{\tau_2(N{-}1)}}\sum_{h=1}^{2(N-1)}\sum_{k_{1}\in \mathbb{Z}} ... \sum_{k_{N_1}\in \mathbb{Z}} \sum_{\hat{k}_{1}\in \mathbb{Z}} ... \sum_{\hat{k}_{N_2}\in \mathbb{Z}}\sum_{\bar{k}_{1}\in \mathbb{Z}} ... \sum_{\bar{k}_{\overline{N}_1}\in \mathbb{Z}}\sum_{\bar{\hat{k}}_{1}\in \mathbb{Z}} ... \sum_{\bar{\hat{k}}_{\overline{N}_2}\in \mathbb{Z}}\nonumber\\
    &\delta_{\sum_{\ell=1}^{2(N-1)}k_{\ell}-h,0}\delta_{\sum_{\ell=1}^{2(N-1)}\bar{k}_{\ell}-h,0}q^{{1\over 2}\sum_{\ell=1}^{2(N-1)}k_{\ell}^{2}-{h^{2}\over 4(N-1)}}\bar{q}^{{1\over 2}\sum_{\ell=1}^{2(N-1)}\bar{k}_{\ell}^{2}-{h^{2}\over 4(N-1)}}\nonumber\\
    &e^{i\pi\left(\sum_{i=1}^{N_1}k_i - \sum_{\bar{i}=1}^{\overline{N}_1}\bar{k}_{\bar{i}}\right)},
\end{align}
which can be written in a more compact notation as in eq. \eqref{integral i}.

\section{Worldsheet integral at level $N=2$ using $SU(2)$ characters}
\label{appendix characters}
The correction to the mass of the level $N=2$ can also be computed using the expression of the Jacobi $\vartheta$ functions in terms of $SO(2n)$ characters. The modular integral is treated using the $i\varepsilon$-prescription described in Section \ref{i epsilon section}, whereas the worldsheet integral takes a different form, but it can be still computed analytically.\\

As an example, we treat the level $N=2$. The worldsheet integral reads
\begin{align}
    \mathcal{I}_{ws}=\int_{T^2} d^2z\, e^{-4\pi\frac{z_2^2}{\tau_2}}\left(\frac{\vartheta_1(z|\tau) \, \overline{\vartheta_1(z|\tau)}}{\vartheta_1'(0|\tau) \, \overline{\vartheta_1'(0|\tau)}}\right)^2.
\end{align}
Following the compact notation $O_{2n}, V_{2n}, S_{2n}, C_{2n}$, introduced in \cite{Bianchi:1990yu, Bianchi:1990tb} for the characters of the $SO(2n)$ current algebra at level $k=1$ 
\begin{equation}
 O_{2n}= {\vartheta_3^n+\vartheta_4^n\over 2 \eta^n}, \:  V_{2n}= {\vartheta_3^n-\vartheta_4^n\over 2 \eta^n}, \: S_{2n} = {\vartheta_2^n+i^{-n}\vartheta_1^n\over 2 \eta^n}, \: C_{2n}= {\vartheta_2^n-i^{-n}\vartheta_1^n\over 2 \eta^n},
\end{equation}
it can be shown that 
\begin{align}
    \vartheta_1^2(z|\tau)=-\eta^2[S_4(z|\tau)-C_4(z|\tau)]=-\eta^2\left[\chi_0(0|\tau)\chi_{\frac{1}{2}}(z|\tau)-\chi_{\frac{1}{2}}(0|\tau)\chi_0(z|\tau)\right],
\end{align}
where the $\chi_j$s are SU(2) characters at level $k=1$ ($j=0, \frac{1}{2}$):
\begin{align}
    \chi_j(z|\tau)=\frac{1}{\eta}\sum_{n\in\mathbb{Z}} q^{(n+j)^2}\zeta^{2(n+j)},\qquad q=e^{2\pi i\tau},\, \zeta=e^{2\pi iz}.
\end{align}
The resulting integral is thus
\begin{align}
    &\int_{T^2} d^2z\, e^{-4\pi\tau_2 y^2}\frac{|\eta|^4}{|\eta|^{12}}\left|\left[\chi_0(0|\tau)\chi_{\frac{1}{2}}(z|\tau)-\chi_{\frac{1}{2}}(0|\tau)\chi_0(z|\tau)\right]\right|^2=\nonumber\\
    =&\,\int_{T^2} d^2z\, e^{-4\pi\tau_2 y^2}\frac{1}{|\eta|^8}\left[\left|\chi_0(0|\tau)\right|^2\left|\chi_{\frac{1}{2}}(z|\tau)\right|^2+\left|\chi_{\frac{1}{2}}(0|\tau)\right|^2\left|\chi_0(z|\tau)\right|^2\right. + \nonumber\\
    & \qquad\qquad\qquad\qquad\,\,\,\, \left.-2\,\mathrm{Re}\left(\chi_0(0|\tau)\overline{\chi_{\frac{1}{2}}}(0|\tau)\chi_{\frac{1}{2}}(z|\tau)\overline{\chi_0(z|\tau)}\right)\right].
\end{align}
The third term yields a vanishing result because the weights of the two representations do not match.\\

As for the $|\chi_0(z|\tau)|^2$ terms, we observe that
\begin{align}
    &\int_{T^2} d^2z\, e^{-4\pi\tau_2 y^2}\sum_{n,\,\bar{n}\in \mathbb{Z}}q^{n^2}\bar{q}^{\bar{n}^2}e^{4\pi iz_1(n-\bar{n})}e^{-4\pi z_2(n+\bar{n})}=\nonumber\\
    =&\,\int_0^{\tau_2} dz_2\, e^{-4\pi\tau_2 y^2} \sum_{n\in\mathbb{Z}}(q\bar{q})^{n^2}e^{-8\pi\tau_2 yn}=\sum_{n\in \mathbb{Z}}\tau_2\int_0^1 dy\,e^{-4\pi\tau_2(y^2+2ny+n^2)}=\frac{\sqrt{\tau_2}}{2}.
\end{align}

The $\left|\chi_{\frac{1}{2}}(z|\tau)\right|^2$ term can be treated in a similar way:
\begin{align}
    &\int_{T^2} d^2z\,e^{-4\pi\tau_2 y^2}\sum_{n,\,\bar{n}\in \mathbb{Z}}q^{(n+\frac{1}{2})^2}\bar{q}^{(\bar{n}+\frac{1}{2})^2}e^{4\pi iz_1(n+\frac{1}{2}-\bar{n}-\frac{1}{2})}e^{-4\pi z_2(n+\frac{1}{2}+\bar{n}+\frac{1}{2})}=\nonumber\\
    =&\,\int_0^{\tau_2} dz_2\, e^{-4\pi\tau_2 y^2} \sum_{n\in\mathbb{Z}}(q\bar{q})^{(n+\frac{1}{2})^2}e^{-8\pi\tau_2 y(n+\frac{1}{2})} = \nonumber\\
    =&\, \sum_{n\in \mathbb{Z}}\tau_2\int_0^1 dy\,e^{-4\pi\tau_2[y^2+2y(n+\frac{1}{2})+(n+\frac{1}{2})^2]}=\frac{\sqrt{\tau_2}}{2}.
\end{align}

Therefore, the worldsheet integral produces the expression
\begin{align}
    \frac{\sqrt{\tau_2}}{2|\eta|^{10}}(|\chi_0|^2+|\chi_{\frac{1}{2}}|^2),
\end{align}
so that the integral over the fundamental region $\mathcal{F}$ of modular parameters becomes:
\begin{align}
    \mathcal{A}^{(2)}=&\int_{\mathcal{F}} \frac{d^2\tau}{\tau_2}\frac{1}{\tau_2^5}\tau_2\frac{\sqrt{\tau_2}}{2|\eta|^{10}}(|\chi_0|^2+|\chi_{\frac{1}{2}}|^2)=\nonumber\\
    =&\,\int_{\mathcal{F}} \frac{d^2\tau}{\tau_2^{\frac{9}{2}}}\frac{1}{2|\eta|^{10}}\left(\left|\chi_0\right|^2+\left|\chi_{\frac{1}{2}}\right |^2\right),
\end{align}
where the additional factor $\tau_2$ comes from the first integration of the cyclic worldsheet coordinate $w$ on the torus. As a consistency check, notice that the integrand is modular invariant.\\
At $z=0$, these characters admit the series expansions
\begin{align}
    &\chi_0(0|\tau) = q^{-\frac{1}{24}}\left[1 + 3q + 4 q^2 + 7 q^3 + 13 q^4 + 
 19 q^5 + 29 q^6 + 43 q^7 + \mathcal{O}(q^8)\right],\\
 &\chi_{\frac{1}{2}}(0|\tau) = 2q^{\frac{5}{24}}\left[1 + q + 3q^2 + 4q^3 + 
 7q^4 + 10q^5 + 17q^6 + 23q^7 + \mathcal{O}(q^8)\right].
\end{align}

When evaluated with the $i\varepsilon$-prescription, this integral yields the same result as the one in Section \ref{n=2 section}, thus serving as a further confirmation for our approach.

\section{A generic integral at level $N$ with lattice sums}
\label{appendix lattice sums}
The worldsheet integral \eqref{general worldsheet integral} has been computed by means of the defining expressions of $\vartheta_1$ and $\vartheta_2$. In this appendix, we present an alternative analytical derivation that relies instead on the decomposition of the Jacobi $\vartheta_s$ functions into $SU(2n)$ and $U(1)_{2n}$ lattice sums.\\

We start from the integral
\begin{align}
    {\cal I}_{N_1,\,\overline{N}_1,\,N_2,\,\overline{N}_2}=\int_{T^2} d^2z\,e^{-4\pi(N-1)\tau_2 y^2}\vartheta_1(z|\tau)^{2N_1}\overline{\vartheta_1(z|\tau)}^{2\overline{N}_1}\vartheta_2(z|\tau)^{2N_2}\overline{\vartheta_2(z|\tau)}^{2\overline{N}_2},
\end{align}
with the constraint $N_1+N_2=N-1=\overline{N}_1+\overline{N}_2$. In general, $N_1\ge 1$ and $\overline{N}_1\ge 1$, while $N_2$ and $\overline{N}_2$ can be zero. The $\vartheta_1$ and $\vartheta_2$ functions can be written as
\begin{align}
    \vartheta_a^{2n}=\left\{\begin{aligned}
        &(-)^n\sum_{r}(-)^r\Lambda^{SU(2n)}_r\xi^{(2n)}_{n-r}\quad\text{if }a=1\\
        &\sum_{r}\Lambda^{SU(2n)}_r\xi^{(2n)}_{n-r}\qquad\,\,\,\,\,\text{if }a=2
    \end{aligned}\right.,
\end{align}
where the $SU(2n)$ and $U(1)_{2n}$ lattice sums are defined as
\begin{align}
  &\Lambda_r^{SU(2n)}=\sum_{m\in\mathbb{Z}^{2n},\,\sum_i m_i=r} q^{\frac{1}{2}\sum_i \left(m_i-\frac{r}{2n}\right)^2},\\
  &\xi^{(2n)}_{n-r}=\sum_{l\in\mathbb{Z}}q^{\frac{1}{2}2nQ^2}e^{2\pi iz2nQ}, \qquad Q:=l+\frac{n-r}{2n}.
  \label{lattice sums definition}
\end{align}

%{\color{red}{MF: I am confused.
%
%et's consider just $\vartheta_1(z,\tau)$. From the formula above it seems that there are three sums... the formula I find and I numerically checked  is 
%$$
%\vartheta_1^{2N}=\sum_{r\in \mathbb{Z} }(-)^r C_r^{(N)}%(q)\,e^{2\pi i r z} q^{r^2\over 4 N}$$
%with
%$$
%C_r^{(N)}(q)=\sum_{m_1\in \mathbb{Z}}...\sum_{m_{2N}\in \mathbb{Z}}\delta_{\sum_{j=1}^{2n}m_j,r-N}\,q^{{1\over 2}\sum_{j=1}^{2n}(m_j+{N-r\over 2N})^2}
%$$
%Can we use this expression to compute the general integral ??
%}}

For the purposes of the worldsheet integral, we can omit the $SU(2n)$ lattice sums (which depend only on the modular parameter $\tau$) and focus on the $U(1)_{2n}$ factors. Therefore, we will consider the integral
\begin{align}
    I_{N_1,\,\overline{N}_1,\,N_2,\,\overline{N}_2} = \int_0^1 dy \int_0^1 dx e^{-4\pi\tau_2(N-1)y^2} &\,\sum_{l_{1,2}\in\mathbb{Z}} q^{N_1Q_1^2 + N_2Q_2^2}\, e^{4i\pi z(N_1Q_1 + N_2Q_2)}\nonumber\\
    &\,\sum_{\bar{l}_{1,2}\in\mathbb{Z}} \bar{q}^{\overline{N}_1\overline{Q}_1^2 + \overline{N}_2\overline{Q}_2^2}\, e^{-4i\pi \bar{z}(\overline{N}_1\overline{Q}_1 + \overline{N}_2\overline{Q}_2)}.
\end{align}
The integral over $x$ results in the condition
\begin{equation}
N_1 Q_1+N_2 Q_2=:L=\overline{L}:=\overline{N}_1 \overline{Q}_1+\overline{N}_2 \overline{Q}_2.
\label{constraint on NQ}
\end{equation}
All the terms proportional to $\tau_1 y$ cancel out
as a consequence of the constraint \eqref{constraint on NQ} and the level-matching constraint $N_1+N_2=\overline{N}_1+\overline{N}_2$. Considering only the terms that depend on $y$, the exponent of the integrand is thus
\begin{align}
    &-4\pi\tau_2 (N-1)y^2-2\pi\tau_2 y(2N_1Q_1+2\overline{N}_1\overline{Q}_1+2N_2Q_2+2\overline{N}_2\overline{Q}_2)=\nonumber\\
    =&\,-4\pi\tau_2(N-1)\left[y^2+2y\frac{N_1Q_1+N_2Q_2}{N-1}+\left(\frac{N_1Q_1+N_2Q_2}{N-1}\right)^2\right]+\nonumber\\
    &\,+4\pi\tau_2\frac{(N_1Q_1+N_2Q_2)^2}{N-1}\equiv\nonumber\\
    \equiv&\,-4\pi\tau_2(N-1)\left[y^2+2y\frac{L}{N-1}+\left(\frac{L}{N-1}\right)^2\right]+4\pi\tau_2\frac{L^2}{N-1}=\nonumber\\
    =&\,-4\pi\tau_2(N-1)\left(y+\frac{L}{N-1}\right)^2+4\pi\tau_2\frac{L^2}{N-1},
    \label{exponent 1}
\end{align}
where we have completed the square to reconstruct a Gaussian integrand in $y$.\\

Assuming $N_1N_2\neq 0$, from eq. \eqref{lattice sums definition} we have four charges\footnote{We will discuss the cases in which $N_2=0$ and/or $\overline{N}_2 =0$ later on.} (two in the holomorphic and two in the anti-holomorphic sectors), namely
\begin{equation}
Q_1:=l_1 + {1\over 2}-{r_1\over 2 N_1}\,,\quad \overline{Q}_1:=\bar{l}_1 + {1\over 2}-{\bar{r}_1\over 2 \overline{N}_1},
\end{equation}
\begin{equation}
Q_2:=l_2 + {1\over 2}-{r_2\over 2 N_2}\,,\quad \overline{Q}_2:=\bar{l}_2 + {1\over 2}-{\bar{r}_2\over 2 \overline{N}_2},
\end{equation}
where the condition \eqref{constraint on NQ} enforces a further constraint
\begin{equation}
    r_1 + r_2 = \bar{r}_1+\bar{r}_2\,\mod2.
    \label{constraint on r1+r2}
\end{equation}
In order to treat the $y$-independent terms of the exponent, it is useful to switch to a parametrization
\begin{align}
    \left\{\begin{aligned}
        &Q_1={L\over N_1+N_2}+{\delta\over N_1}\,,\qquad \overline{Q}_1={\overline{L}\over \overline{N}_1+\overline{N}_2}+{\bar{\delta}\over \overline{N}_1} = {L\over \overline{N}_1+\overline{N}_2}+{\bar{\delta}\over \overline{N}_1}\\
        &Q_2={L\over N_1+N_2}-{\delta\over N_2}\,,\qquad \overline{Q}_2={\overline{L}\over \overline{N}_1+\overline{N}_2}-{\bar{\delta}\over \overline{N}_2} = {L\over \overline{N}_1+\overline{N}_2}-{\bar{\delta}\over \overline{N}_2},
    \end{aligned}\right.
    \label{parametrization p1}
\end{align}
where we have already used the constraint \eqref{constraint on NQ}.\\

Using the parametrization \eqref{parametrization p1}, we move to combine the last term in the expression \eqref{exponent 1} with the exponents of $q$ and $\bar{q}$ appearing in $\xi$ and $\overline{\xi}$:
\begin{align}
    &2\pi i\tau N_1Q_1^2+2\pi i\tau N_2Q_2^2-2\pi i\overline{\tau}\overline{N}_1\overline{Q}_1^2-2\pi i\overline{\tau}\overline{N}_2\overline{Q}_2^2+4\pi\tau_2\frac{L^2}{N{-}1}=\nonumber\\
    =&\,2\pi i\tau\left[N_1\left(\frac{L}{N{-}1}+\frac{\delta}{N_1}\right)^2+N_2\left(\frac{L}{N{-}1}-\frac{\delta}{N_2}\right)^2\right]+\nonumber\\
    &-2\pi i\overline{\tau}\left[\overline{N}_1\left(\frac{L}{N{-}1}+\frac{\overline{\delta}}{\overline{N}_1}\right)^2+\overline{N}_2\left(\frac{L}{N{-}1}-\frac{\overline{\delta}}{\overline{N}_2}\right)^2\right]+4\pi\tau_2\frac{L^2}{N{-}1}=\nonumber\\
    =&\,2\pi i\tau\left[N_1\frac{L^2}{(N{-}1)^2}+2\frac{L\delta}{N{-}1}+\frac{\delta^2}{N_1}+N_2\frac{L^2}{(N{-}1)^2}-2\frac{L\delta}{N{-}1}+\frac{\delta^2}{N_2}\right]+\nonumber\\
    &\,-2\pi i\overline{\tau}\left[\overline{N}_1\frac{L^2}{(N{-}1)^2}+2\frac{L\overline{\delta}}{N{-}1}+\frac{\overline{\delta}^2}{\overline{N}_1}+\overline{N}_2\frac{L^2}{(N{-}1)^2}-2\frac{L\overline{\delta}}{N{-}1}+\frac{\overline{\delta}^2}{\overline{N}_2}\right]+4\pi\tau_2\frac{L^2}{N{-}1}=\nonumber\\
    =&\,2\pi i\tau(N_1{+}N_2)\frac{L^2}{(N{-}1)^2}+2\pi i\tau\left(\frac{1}{N_1}+\frac{1}{N_2}\right)\delta^2+\nonumber\\
    &-2\pi i\overline{\tau}(\overline{N}_1{+}\overline{N}_2)\frac{L^2}{(N{-}1)^2}-2\pi i\overline{\tau}\left(\frac{1}{\overline{N}_1}+\frac{1}{\overline{N}_2}\right)\overline{\delta}^2+4\pi\tau_2\frac{L^2}{N{-}1}=\nonumber\\
    =&\,2\pi i\tau\left(\frac{1}{N_1}+\frac{1}{N_2}\right)\delta^2-2\pi i\overline{\tau}\left(\frac{1}{\overline{N}_1}+\frac{1}{\overline{N}_2}\right)\overline{\delta}^2,
\end{align}
thus yielding the generic term
\begin{align}
    q^{\left(\frac{1}{N_1}+\frac{1}{N_2}\right)\delta^2}\overline{q}^{\left(\frac{1}{\overline{N}_1}+\frac{1}{\overline{N}_2}\right)\overline{\delta}^2},
\end{align}
%Overall, the integrand will be in the form
%\begin{equation}
%\sum q^{\left(\frac{1}{N_1}+\frac{1}{N_2}\right)\delta^2}\overline{q}^{\left(\frac{1}{\overline{N}_1}+\frac{1}{\overline{N}_2}\right)\overline{\delta}^2}\, e^{-4\pi(N_1+N_2)\tau_2\left(y+{L\over N_1+N_2}\right)^2}
%\end{equation}
where $\delta$ can be expressed in terms of the indices appearing in the definitions \eqref{lattice sums definition} as
\begin{equation}
\left(\frac{1}{N_1}+\frac{1}{N_2}\right)\delta=Q_1-Q_2=l_1-l_2-\frac{r_1}{2N_1}+\frac{r_2}{2N_2}.
\end{equation}

Our next objective is to express $L$, $\delta$ and $\bar{\delta}$ in such a way as to simplify the sums to compute. To this purpose, we introduce the new parametrization
\begin{align}
    \left\{\begin{aligned}
        &N_1 l_1 + N_2 l_2=(N_1+N_2)\ell + \lambda\\
        &\overline{N}_1 \bar{l}_1 + \overline{N}_2 \bar{l}_2=(\overline{N}_1+\overline{N}_2)\bar{l} + \bar{\lambda}\\
        &l_1-l_2=m\\
        &\bar{l}_1-\bar{l}_2=\overline{m}
    \end{aligned}\right.,\qquad \ell,\,\bar{\ell},\lambda,\,\bar{\lambda},\,m,\,\overline{m}\in\mathbb{Z},
    \label{parametrization p2}
\end{align}
so that we can express $\delta$ and $\bar{\delta}$ in terms of integer indices:
\begin{equation}
\left\{\begin{aligned}
    &\delta={1\over N_1+N_2}\left[ N_1 N_2 m + {1\over 2}(N_1 r_2-N_2r_1) \right]\\
    &\bar{\delta}={1\over \overline{N}_1+\overline{N}_2}\left[ \overline{N}_1 \overline{N}_2 \overline{m} + {1\over 2}(\overline{N}_1\bar{r}_2-\overline{N}_2\bar{r}_1) \right]\end{aligned}\right..
\end{equation}
Furthermore, the original indices $l_{1,2}$ and $\bar{l}_{1,2}$ can be rewritten as
\begin{equation}
\left\{\begin{aligned}
    &l_1=\ell + {\lambda+N_2 m \over N_1+N_2}\\
    &l_2=\ell +{\lambda - N_1 m\over N_1+N_2}\\
    &\bar{l}_1=\bar{\ell} + {\bar{\lambda}+\overline{N}_2 \overline{m} \over \overline{N}_1+\overline{N}_2}\\
    &\bar{l}_2=\bar{\ell} +{\bar{\lambda} - \overline{N}_1 \overline{m}\over \overline{N}_1+\overline{N}_2}\end{aligned}\right..
\end{equation}

As far as the $\delta$-independent part of the integrand is concerned, we can proceed as follows. We start by writing
\begin{align}
    &L=N_1Q_1+N_2Q_2=N_1 l_1+N_2 l_2+\frac{N_1+N_2}{2}-\frac{r_1+r_2}{2}\\[10pt]
    &\Rightarrow\qquad \frac{L}{N_1+N_2}=\ell+\frac{\lambda}{N_1+N_2}+\frac{1}{2}-\frac{r_1+r_2}{2(N_1+N_2)}.
\end{align}
The sum over $\ell$ can be used to reconstruct the Gaussian integral over $y$, yielding
\begin{flalign}
    \sum_{\ell\in\mathbb{Z}}\int_0^1 dy\,e^{-4\pi\tau_2(N-1)\left(y+\ell+\frac{\lambda}{N-1}+\frac{1}{2}-\frac{r_1+r_2}{2(N-1)}\right)^2}=\sqrt{\frac{1}{4\tau_2(N-1)}}.
\end{flalign}

Moreover, since $l_{1,2}$, $\bar{l}_{1,2}$, $\ell$ and $\bar{\ell}$ are all integers, the parametrization \ref{parametrization p2} also introduces the projectors
\begin{flalign}
    \mathrlap{\left\{\begin{aligned}
    &\frac{1}{N_1+N_2}\sum_{k_1=0}^{N_1+N_2-1}e^{2\pi i {k_1\over N_1+N_2}(\lambda + N_2 m)}\,,\,\,\, \frac{1}{N_1+N_2}\sum_{k_2=0}^{N_1+N_2-1}e^{2\pi i {k_2\over N_1+N_2}(\lambda - N_1 m)}\\
    &\frac{1}{\overline{N}_1+\overline{N}_2}\sum_{\bar{k}_1=0}^{\overline{N}_1+\overline{N}_2-1}e^{-2\pi i {\bar{k}_1\over \overline{N}_1+\overline{N}_2}(\bar{\lambda} + \overline{N}_2 \overline{m})}\,,\,\,\, \frac{1}{\overline{N}_1+\overline{N}_2}\sum_{\bar{k}_2=0}^{\overline{N}_1+\overline{N}_2-1}e^{-2\pi i {\bar{k}_2\over \overline{N}_1+\overline{N}_2}(\bar{\lambda} - \overline{N}_1 \overline{m})}\end{aligned}\right..}&&
    \label{projectors p2}
\end{flalign}
The constraint \ref{constraint on NQ} now reads
\begin{align}
(N_1+N_2)\ell+\lambda+\frac{1}{2}-\frac{r_1+r_2}{2}=(\overline{N}_1+\overline{N}_2)\bar{\ell}+\bar{\lambda}+\frac{1}{2}-\frac{\bar{r}_1+\bar{r}_2}{2}.
\end{align}
Therefore, we have
\begin{align}
\bar{\ell}=\ell+\frac{\lambda-\bar{\lambda}-\frac{r_1+r_2-\bar{r}_1-\bar{r}_2}{2}}{N_1+N_2},
\end{align}
which we have already implicitly used to reconstruct the Gaussian integral. Since both $\ell$ and $\bar{\ell}$ are integers, the numerator $\lambda-\bar{\lambda}-\frac{r_1+r_2-\bar{r}_1-\bar{r}_2}{2}$ must be a multiple of $N_1+N_2$, thus resulting in the condition
\begin{align}
\bar{\lambda}=\lambda+ {\bar{r}_1+\bar{r}_2-r_1-r_2\over 2}+p(N_1+N_2),\qquad p\in\mathbb{Z}.
\label{lambda lambdabar condition}
\end{align}
Notice that $\lambda$ and $\bar{\lambda}$ are integers as well, thus enforcing the equality $\bar{r}_1+\bar{r}_2=(r_1+r_2)\mod2$, which, however, is already deduced from eq. \ref{constraint on NQ}. Defining $R=r_1+r_2$ and $\bar{R}=\bar{r}_1+\bar{r}_2$, we can rewrite the condition \eqref{lambda lambdabar condition} as
\begin{equation}
\bar{\lambda}=\lambda+\frac{\overline{R}-R}{2}+p(N_1+N_2).
\end{equation}
If we plug this expression into the projectors \eqref{projectors p2} and reinterpret the $\lambda$-dependent part from the point of view of the summation over $\lambda$, we observe that
\begin{align}
&\sum_{\lambda\in\mathbb{Z}}\exp\left\{\frac{2\pi i}{N_1+N_2}\left[k_1(\lambda+N_2m)+k_2(\lambda-N_1m)-\bar{k_1}\left(\lambda+\frac{\overline{R}-R}{2}+\overline{N}_2\overline{m}\right)+\right.\right.\nonumber\\
&\qquad\qquad\qquad\qquad\qquad\left.\left.-\bar{k}_2\left(\lambda+\frac{\overline{R}-R}{2}-\overline{N}_1\overline{m}\right)\right]\right\}\nonumber\\[10pt]
&\qquad\qquad\qquad\Rightarrow\quad k_1+k_2-\bar{k}_1-\bar{k}_2=p(N_1+N_2),\quad p\in\mathbb{Z}.
\end{align}
The sum over $\bar{\lambda}$ cancels one power of $N-1$, while reinterpreting the sum over $\lambda$ as a projector reabsorbs one more such factor $\frac{1}{N-1}$. The rest of the exponent reads
\begin{align}
k_1N_2m-k_2N_1m-\bar{k}_1\frac{\overline{R}-R}{2}-\bar{k}_1\overline{N}_2\overline{m}-\bar{k}_2\frac{\overline{R}-R}{2}+\bar{k}_2\overline{N}_1\overline{m}.
\end{align}

We now set $\bar{k}_2=k_1+k_2-\bar{k}_1-p(N_1+N_2)$. The original constraint $N_1Q_1+N_2Q_2=\overline{N}_1\overline{Q}_1+\overline{N}_2\overline{Q}_2$ forces $\frac{\overline{R}-R}{2}$ to be an integer. Since also all the other factors are integers, all the terms that are proportional to $p(N_1+N_2)$ drop from the exponent. The sum over $\bar{k}_1$ drops in this computation and thus cancels the remaining power of $N-1$, so that in the end the exponent becomes
\begin{align}
    -(k_1+k_2)\left(N_1m+\frac{\overline{R}-R}{2}-\overline{N}_1\overline{m}\right).
\end{align}
Therefore, considering all the projectors involved, the complete form of the integrand becomes
\begin{align}
I_{N_1,\overline{N}_1,N_2,\overline{N}_2}=&\,(-)^{N_1+\overline{N}_1}\frac{\sqrt{\tau_2(N-1)}}{2}\sum_{r_1,r_2,\bar{r}_1,\bar{r}_2}^{0,\, 2N_i-1}(-)^{r_1+\bar{r}_1}\Lambda_{r_1}^{SU(2N_1)}\bar{\Lambda}_{\bar{r}_1}^{SU(2\overline{N}_1)}\Lambda_{r_2}^{SU(2N_2)}\bar{\Lambda}_{\bar{r}_2}^{SU(2\overline{N}_2)}\nonumber\\
&\,\sum_{\varepsilon=1}^2 e^{\pi i\varepsilon (\bar{r}_1+\bar{r}_2-r_1-r_2)}\sum_{m,\overline{m}\in\mathbb{Z}}q^{\left(\frac{1}{N_1}+\frac{1}{N_2}\right)\delta^2} \bar{q}^{\left(\frac{1}{\overline{N}_1}+\frac{1}{\overline{N}_2}\right)\bar{\delta}^2}\nonumber\\
&\,\sum_{k_1=1}^{N-1} e^{-\frac{2\pi i}{N-1}k_1\left(N_1 m+\frac{\bar{r}_1+\bar{r}_2-r_1-r_2}{2}-\overline{N}_1\overline{m}\right)}.
\end{align}

Similar expressions can be obtained when some of the $\vartheta_i$ factors are missing in the integral. As a result, we find
\begin{align}
    I_{N_1,\overline{N}_1,0,0}&:= \int_{T^2} d^2z\,e^{-4\pi(N-1)\tau_2 y^2}\vartheta_1^{2N_1}\overline{\vartheta}_1^{2\overline{N}_1}=\nonumber\\
    &=\frac{1}{2}\sqrt{\frac{\tau_2}{N-1}}\sum_{r_1,\bar{r}_1}^{0,\,2N_i-1}\Lambda_{r_1}^{SU(2N_1)}\bar{\Lambda}_{\bar{r}_1}^{SU(2\overline{N}_1)}\frac{1+(-)^{\bar{r}_1-r_1}}{2},
\end{align}
\begin{align}
    I_{N_1,\overline{N}_1,0,\overline{N}_2}:=& \int_{T^2} d^2z\,e^{-4\pi(N-1)\tau_2 y^2}\vartheta_1^{2N_1}\overline{\vartheta}_1^{2\overline{N}_1}\overline{\vartheta}_2^{2\overline{N}_2}=\nonumber\\
    =&(-)^{N_1+\overline{N}_1}\frac{1}{2}\sqrt{\frac{\tau_2}{N-1}}\sum_{r_1,\bar{r}_1,\bar{r}_2}^{0,\,2N_i-1}\Lambda_{r_1}^{SU(2N_1)}\bar{\Lambda}_{\bar{r}_1}^{SU(2\overline{N}_1)}\bar{\Lambda}_{\bar{r}_2}^{SU(2\overline{N}_2)}\frac{1+(-)^{\bar{r}_1+\bar{r}_2-r_1}}{2}\cdot\nonumber\\
    &\cdot\sum_{\overline{m}\in\mathbb{Z}}\bar{q}^{\left(\frac{1}{\overline{N}_1}+\frac{1}{\overline{N}_2}\right)\bar{\delta}^2}\sum_{\bar{k}_1=1}^{N-1}e^{-\frac{2\pi i \bar{k}_1}{N-1}\left(\frac{\bar{r}_1+\bar{r}_2-r_1}{2}+\overline{N}_2\overline{m}\right)}
\end{align}
and
\begin{align}
    I_{N_1,\overline{N}_1,N_2,0}\equiv& \int_{T^2} d^2z\,e^{-4\pi(N-1)\tau_2 y^2}\vartheta_1^{2N_1}\overline{\vartheta}_1^{2\overline{N}_1}\vartheta_2^{2N_2}=\nonumber\\
    =&(-)^{N_1+\overline{N}_1}\frac{1}{2}\sqrt{\frac{\tau_2}{N-1}}\sum_{r_1,r2,\bar{r}_1}^{0,\,2N_i-1}\Lambda_{r_1}^{SU(2N_1)}\bar{\Lambda}_{\bar{r}_1}^{SU(2\overline{N}_1)}\Lambda_{r_2}^{SU(2N_2)}\frac{1+(-)^{\bar{r}_1-r_1-r_2}}{2}\cdot\nonumber\\
    &\cdot\sum_{m\in\mathbb{Z}}q^{\left(\frac{1}{N_1}+\frac{1}{N_2}\right)\delta^2}\sum_{k_1=1}^{N-1}e^{\frac{2\pi i k_1}{N-1}\left(N_2 m+\frac{r_1+r_2-\bar{r}_1}{2}\right)}.
\end{align}

\bibliographystyle{JHEP}
\bibliography{stringbib}

\end{appendix}

\end{document}